\documentclass[preprint,onecolumn,superscriptaddress,amsmath,amssymb,aps,floatfix,nofootinbib,longbibliography]{revtex4-1}

\usepackage{multirow}
\usepackage[pdftex]{graphicx}
\usepackage{bm}
\usepackage{color}
\usepackage{times}
\usepackage{amsmath}
\usepackage{amssymb}
\usepackage{mathtools}
\usepackage{gensymb} 
\newcommand{\defeq}{\vcentcolon=}
\usepackage{tablefootnote}
\usepackage{textcomp}
 \usepackage{dsfont}\usepackage{subfig}
 \usepackage{float}
\usepackage{caption}
 \usepackage{ragged2e}
\usepackage[utf8]{inputenc}
\usepackage{natbib}
\usepackage{wrapfig}
\usepackage{lipsum}
\usepackage{xcolor}
\usepackage[normalem]{ulem}
\usepackage{amsfonts}
\usepackage{tikz, todonotes}
\usetikzlibrary{shapes,shadows,arrows}

\usepackage{url}
\usepackage[english]{babel}

\usepackage{booktabs}
\usepackage[colorlinks=true, allcolors=blue]{hyperref}
\usepackage{stackrel} % Stack operators
\usepackage{polynom} % Divide polynomials
\polyset{style=A, div=:, vars=x}

\usepackage[normalem]{ulem}
\usepackage{multirow}

\usepackage{colortbl}

\begin{document}
%\linenumbers

\title{Development and experimental validation of a mathematical model for fluoride-removal filters comprising chemically treated mineral rich carbon}

\author{Lucy C. Auton}
\affiliation{Centre de Recerca Matem\`{a}tica,  Campus de Bellaterra, Barcelona, Spain} 

% Edifici C, Facultat de Ciències
% 08193 Bellaterra, 
%Barcelona, 

\author{Shanmuk S. Ravuru}
\affiliation{Dept. Chem. \& Mat. Eng., University of Alberta, Canada}

\author{Sirshendu De}
%\email{christopher.macminn@eng.ox.ac.uk}
\affiliation{Dept. Chem. Eng., Indian Institute of Technology Kharagpur, India}

\author{Tim Myers}
\affiliation{Centre de Recerca Matem\`{a}tica,  Campus de Bellaterra, Barcelona, Spain}

\author{Abel Valverde}
\affiliation{Dept. Chem. Eng, Universitat Polit\`{e}cnica de Catalunya, Escola Superior \\ d’Enginyeries Industrial,\ Aeroespacial i Audiovisual de Terrassa, Spain}

%

%\date{\today}
%\linenumbers
\vspace{-1cm}

\begin{abstract}
Excessive fluoride intake can lead to dental and skeletal fluorosis, among other health issues. Naturally occurring fluoride and industrial runoff can result in concentrations far exceeding the World Health Organization’s recommended limits in water supplies. In this study, we derive a  model incorporating the dominant mechanisms governing fluoride removal from drinking water using the two adsorbents mineral-rich carbon (MRC) and chemically treated mineral-rich carbon (TMRC). Using both new and previously published experimental data, we validate the model for MRC, TMRC, and their mixture, {using both batch and column data.}
Despite the filters containing  approximately 40:1 MRC:TMRC ratio by mass, we find that TMRC dominates fluoride removal, while MRC contributes at early and late times. The full column model, which uses parameters from isotherm  batch studies, achieves excellent agreement with experimental breakthrough data across varying inlet concentrations and flow rates (R$^2>0.991$, SSE$<0.0632$). Motivated by this, we propose a reduced model based solely on TMRC adsorption, with a single  fitting parameter, which still performs well across all  breakthrough curves (R$^2 > 0.983$, SSE $<0.117$). {The simplicity of this model means that it is straightforward and inexpensive to work with numerically.} 
In both models, batch and column behaviours are reconciled and, for the case of breakthrough curves with varying inlet concentrations, a set of globally optimised parameters is found. The strong agreement with experimental data supports the model's robustness and reinforces the physical interpretability of its parameters. These models for MRC and TMRC provide a foundation for filter optimisation and future efforts aimed at improving fluoride removal in resource-limited settings.
\end{abstract}

\textit{Corresponding author: tmyer@crm.cat }

% \begin{keyword}{Fluoride removal, Mineral rich carbon, Adsorption, Column sorption, Batch adsorption, Mathematical modelling}
% \end{keyword}

\maketitle

%\editmarkerA{}

%%%%%%%%%%%%%%%%%%%%%%%%%%%%%%%%%%%%%%%%%%%%%%%%%%%%%%
\section{Introduction}
\label{s:intro}
%%%%%%%%%%%%%%%%%%%%%%%%%%%%%%%%%%%%%%%%%%%%%%%%%%%%%%

%
%%%%%%%%%%%%%%%%%%%%%%%%%%%%%%%%%%%%%%%%%%%%%%%%%%%%%%
{Fluoridation of drinking water is commonplace with an estimated 5.7\% of the global population consuming water with added fluoride~\cite{cheng2007adding}. Countries including the United States, Canada, Ireland, Chile and Australia fluoridate over 50\% of their water; for example in
the United States nearly 60\% of the population receiving fluoridated water~\cite{cheng2007adding}.} The primary motivation for water fluoridation is the prevention of dental caries, as small amounts of fluoride (F$^-$) are essential for healthy teeth and bone development. For this reason, many toothpastes also contain fluoride~\cite{kaminsky1990fluoride}.
However, excessive fluoride intake can lead to adverse health effects, including dental and skeletal fluorosis, and has been linked to other conditions such as cancer, gastrointestinal distress, and neurological damage~\cite{bharti2017fluoride,kaminsky1990fluoride}.  Fluoride is naturally present in certain rocks and soils and can leach into groundwater; it can also enter the environment through industrial runoff and atmospheric aerosols, impacting ecosystems~\cite{ozsvath2009fluoride}. {In  India, approximately 62 million people are estimated to consume water containing more than  the World Health Organization’s (WHO) recommended maximum concentration of 1.5~mg/l ($\sim 7.89\times10^{-5}\ $mol/l) of F$^-$~\cite{world2019preventing}. This has led to  both skeletal and dental fluorosis becoming endemic in India in at least 20 states. }

\begin{figure}[htb]
    \centering  
    \includegraphics[width=1\textwidth]{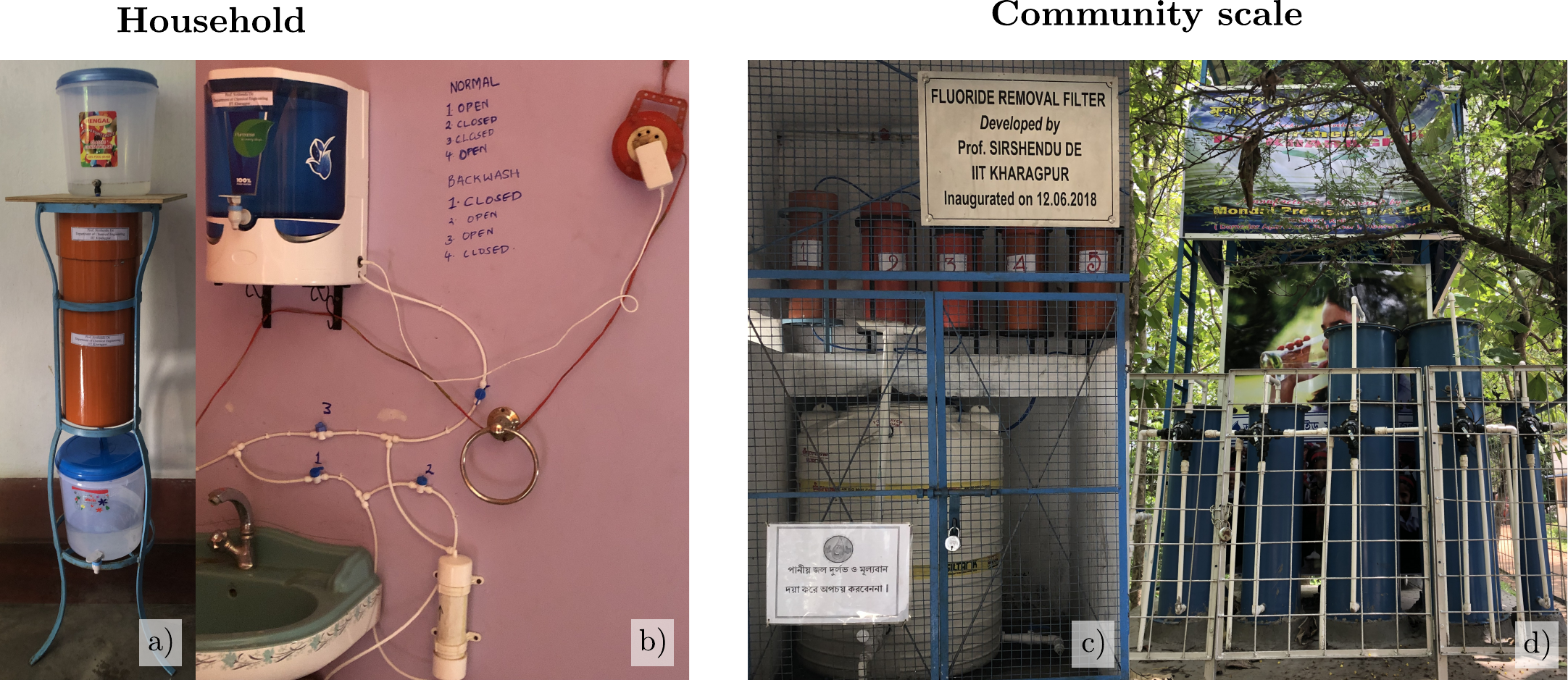}
    \caption{\small Photos of four different fluoride-removal filters in situ in West Bengal: a) A prototype for a household filter. b) A commercialised version of the household filter c) A  prototype for a community-scale filter, situated in a school d) A solar powered, community-scale, commercial filter in a rural village.
    \label{filter_photos}}
\end{figure}

{In this paper, we investigate fluoride adsorption filters composed of mineral-rich carbon (MRC) and chemically treated mineral-rich carbon (TMRC) which are currently being sold to the Indian government and distributed across West Bengal. } Figure~\ref{filter_photos} shows examples of prototype and commercial filters at both household and community scales. The commercial community scale filter (Figure~\ref{filter_photos}d), is powered by solar panels. The water filters through various columns, including the mixture of MRC and TMRC and a column of activated carbon which is known for its antibacterial properties. Thus, as well  as removing fluoride, the filter additionally removes~iron and reduces the bacteria present in the water.  Fluoride levels in household wells in the region vary widely—even wells located within a few hundred meters of each other can differ drastically.

Our goal is to understand the chemical processes occurring within filters comprising MRC and TMRC and to develop accurate, predictive models that capture the dominant reactions involved in fluoride removal. Such models enable reliable predictions under varying experimental conditions and facilitate practical applications. This predictive power supports optimisation of filter design, extending operational life and improving cost-efficiency—ultimately contributing to improved access to safe drinking water, a critical goal for sustainable development and public health.

Both TMRC and MRC have proven highly effective in removing fluoride from water. Chatterjee et al.~\cite{chatterjee2018defluoridation} report that TMRC---   referred to therein  as chemically treated carbonised bone meal (CTBM)---     has an adsorption capacity of 150~mg/g, significantly outperforming other bio-based adsorbents such as aluminium-treated activated bamboo charcoal (21.1~mg/g)~\cite{wendimu2017aluminium}. Mineral rich carbon---     referred to as carbonised bone meal or CBM in~\cite{chatterjee2018novel}---    exhibits an adsorption capacity of 14~mg/g. Among 102 adsorbents reviewed by Bhatnagar et al.~\cite{bhatnagar2011fluoride}, only four outperform TMRC: nanomagnesia, CaO nanoparticles, calcined Mg-Al-CO$_3$, and Fe--Al--Ce trimetal oxide.

\citet{chatterjee2018defluoridation} modelled TMRC adsorption using Langmuir, Freundlich, and Dubinin--Radushkevich isotherms combined with a dual-porosity kinetic model for column data, though they failed to achieve a good fit. Similarly, for MRC, they applied the same isotherms but employed the Yoon--Nelson model for column studies~\cite{chatterjee2018novel}. {Mathematically the isotherm derives from the steady-state of the kinetic model, hence the use of dual-porosity or Yoon-Nelson with unrelated isotherms is inconsistent and this has been shown to result in a poor fit to experimental data or alternatively force system constants to vary with operating parameters, see  \cite{myers2024time, myers2023development}.}

Mineral rich carbon, is also known as carbonised bone meal or, equivalently, bone char. 
\citet{medellin2014adsorption,huyen2023bone,alkurdi2019bone}, find that approximately  70\% -- 90\% of bone char (MRC) comprises hydroxyapatite and they  state that it is the hydroxyapatite in the bone char which is predominantly responsible for the uptake of fluoride. 
 \citet{russo2024fluoride} present a   model for fluoride adsorption on hydroxyapatite both for batch and column adsorption using the Adsorption Dynamic Intraparticle Model (ADIM), derived in~\citet{russo2015dynamic} and based on~\citet{do1998adsorption}. This model incorporates mass conservation in bulk flow and intra-particle diffusion through both liquid in the `pores' and the `solid' phase, assuming local equilibrium represented by a Freundlich isotherm. However, several inconsistencies arise raising doubt as to whether the mechanics of the system can be effectively predicted using their  model.

{The present work stems from the preliminary study described in the conference paper of \citet{auton2023mathematical}. In this previous work we presented, but did not derive, batch  models  for fluoride removal by MRC and TMRC and column sorption models for their mixture and showed that they  matched experimental data far better than  Langmuir's model, particularly under physically realistic conditions - the focus was on the comparison with Langmuir's model, with minimal context given, limited analysis and results which address only this comparison.
Here, we give the complete story;  in addition to providing significantly more background information and context than is contained within \cite{auton2023mathematical},  we present a careful  derivation of  the  models for MRC and TMRC based on  their distinct chemical structures, presenting chemical reactions involved in the removal process and justifying our choice of dominant reactions (MRC: \S\ref{MRC batch}, TMRC: \S\ref{TMRC batch}). 

The individual adsorbents are first characterised through batch isotherm and kinetic experiments (MRC: \S\ref{MRC batch comp}, TMRC: \S\ref{TMRC batch comp}), and the resulting models are then combined into a model for column filters comprising a mixture of MRC and TMRC (\S\ref{Column}). 
We have also reduced the number of  batch isotherm fitting parameters by two, resulting in four intrinsic parameters --- as opposed to the six employed in \cite{auton2023mathematical}--- to be fixed by batch isotherm data.  We now do not fit  any quantities that are directly measurable and the model is consistent with all modelling assumptions. 
Further, the present work includes previously unpublished experimental data which investigates the effect of varying flow rate in a column filter; we investigate the adsorption behaviour within the filter and  consider all quantities at the filter outlet, not just the concentration of fluoride as was the only outlet quantity discussed in \cite{auton2023mathematical}.  

The following model shows strong agreement with breakthrough curves from column experiments, accurately capturing system behaviour across a range of inlet concentrations and flow rates (R$^2 > 0.991$, SSE $< 0.0632$). Most notably, it accurately captures changes in breakthrough behaviour as the inlet fluoride concentration is varied, without requiring model refitting.  Column study results confirm that TMRC is the dominant adsorbent, allowing for a reduced version of the model with only a single  fitting parameter (this was not  considered in \cite{auton2023mathematical}).  Despite its simplicity, the reduced model --- with just one fitting parameter --- still closely matches experimental results (R$^2 > 0.983$, SSE $<0.117$), supporting its use as an computationally inexpensive, efficient and robust predictive tool.}

\section{Chemical model derivation and validation with batch experiments}

We consider the filtration of  water contaminated with the passive solute fluoride,~(F$^-$), by the  adsorbents MRC \cite{chatterjee2018novel}, and TMRC \cite{chatterjee2018defluoridation};  MRC  is carbonised mammalian or avian bone meal, while TMRC is a derivative of MRC which has been chemically treated to improve its fluoride adsorption capacity.   The TMRC is made by grinding down MRC and coating it in  Aluminium Hydroxide $\big(\text{Al(OH)}_3\big)$. 
This treatment process makes TMRC more expensive than MRC, since the cost of aluminium significantly  exceeds the cost of  bone meal. Further,  TMRC has an average grain size of 0.1--0.3~mm  which is notably smaller than MRC, which has an average grain size of 0.4--0.6~mm \cite{chatterjee2018defluoridation}: the smaller the grain size the more prone the filter is to clogging. As a result a filter made of pure TMRC would be undesirable.  We consider experimental data from two different types of filter:  batch filters, which are predominantly experimental tools (Figure~\ref{Batch diagram}, left) and column filters, which is the type of filter used in practice (Figure~\ref{Batch diagram}, right).     

\subsection{Materials and methods}\label{MatandMeth}

{  Chatterjee et al. \cite{chatterjee2018removal,chatterjee2018novel,chatterjee2018defluoridation} describe the creation of MRC and TMRC and provide a full analysis of their properties - the experiments presented in the current work follow the  methodology of \cite{chatterjee2018novel, chatterjee2018defluoridation}. 
Two different experimental setups are considered: batch (Figure \ref{Batch diagram}, left) and column (or fixed-bed) adsorption experiments (Figure \ref{Batch diagram}, right). 
 We use the batch experiments to determine the intrinsic properties of the fluid--adsorbate--adsorbent system, while both prototype and commercialised filters are employed as column filters. All experiments are carried out with a fluoride solution in the absence of other competing ions (such as calcium, aluminium, nitrate, sulfate, phosphate or chloride).

\begin{figure}[htb]
    \centering 
    \includegraphics[width=.9\textwidth]{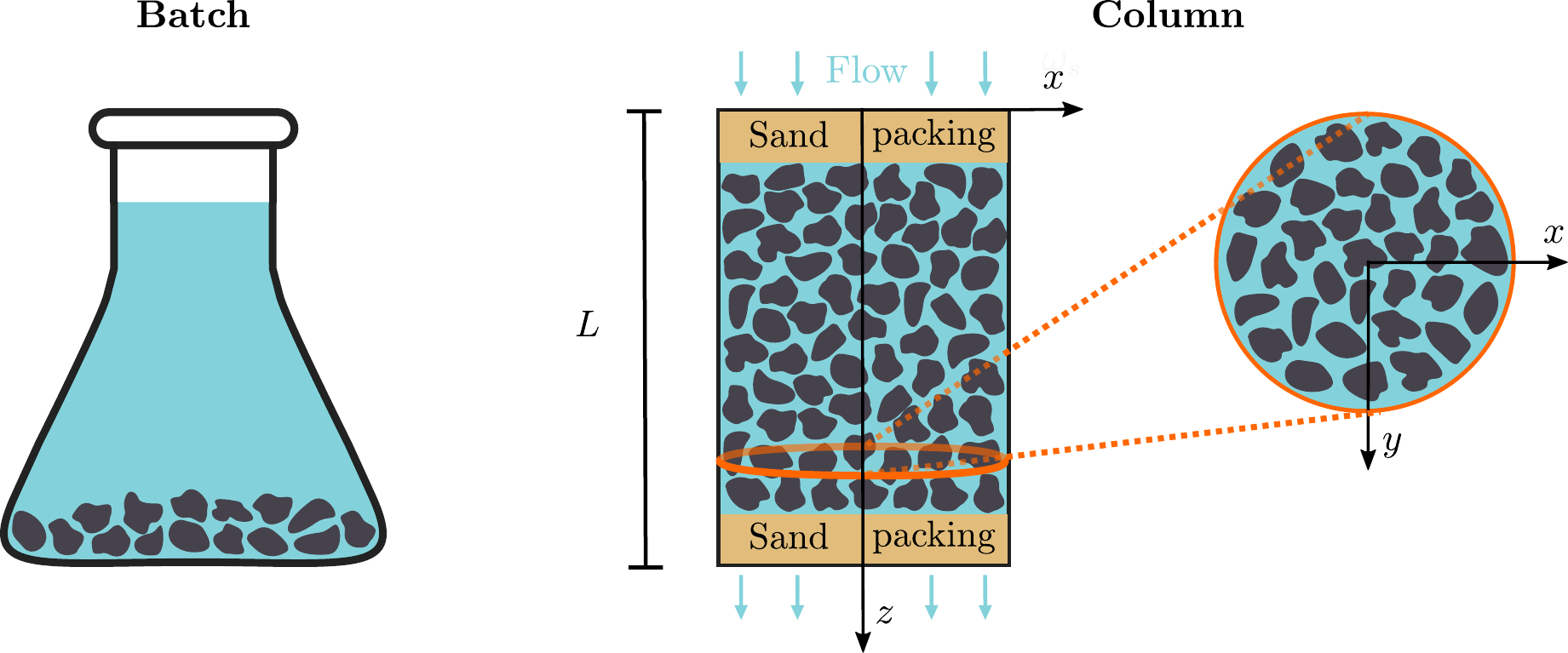}
    \caption{\small %\red{Updated 2/5/25}
   Schematic of batch experiments (left) and column experiments (right).  
    \textbf{\textit{Left}}:  A beaker containing a small amount of adsorbent  surrounded by contaminated fluid; this forms a closed system with no net flow.  
    \textbf{\textit{Right}}: Contaminated fluid flows through a cylinder evenly packed with an adsorbent. The sand at the inlet and outlet of the filter ensures the water enters the filter uniformly. 
    Note that, the porosity of the packing  (and hence fluid fraction)  is much lower   in column than in  batch filters. 
}
    \label{Batch diagram}
\end{figure}

In batch experiments a   contaminated fluid, with a known concentration of fluoride $c_\mathrm{F}$,  is added to a beaker (Figure \ref{Batch diagram}, left)  which contains  a fixed dose (g/l) of adsorbent. The dose determines the ratio of adsorbent to fluoride-contaminated water. This  forms a closed system so that all changes in $c_\mathrm{F}$ are due to removal by the adsorbent.    The beaker is then placed in a shaker which homogenises the adsorbent-fluid mixture.   We use two distinct doses for the batch experiments: for the isotherm experiments 7~g/l of adsorbent is used and only the final equilibrium state is recorded; for the kinetic studies  1~g/l of adsorbent is used and the concentration of fluoride is tracked over time. 
We use experimental data for both the kinetic and isotherm studies for MRC and TMRC, separately. 
%%%%%%%%%%%%%%%%%%%%%%

For the isotherm study, we record only the final concentration --- that is, the beaker remains in the shaker until equilibrium is reached, at which point the adsorbent is filtered from the fluid and the final fluoride concentration is measured yielding a single data point. We repeat this procedure for various initial  concentrations, ranging from $2.5 \times 10^{-4}$ to $0.05$~mol/l (4.75 to 950~mg/l). This results in an isotherm curve which relates the equilibrium adsorbed mass of contaminant per unit mass of adsorbent, $q^\mathrm{e}$, to the corresponding equilibrium fluoride concentration,~$c_\mathrm{F}^\mathrm{e}$.

For the kinetic study, the same beaker–shaker setup is used: for MRC   $c_\mathrm{F}^\mathrm{i} = 5.26 \times 10^{-4}$~mol/l (10~mg/l); for TMRC 
 $c_\mathrm{F}^\mathrm{i} = 2.63 \times 10^{-3}$~mol/l (50~mg/l). We measure the fluoride concentration  at various times by extracting small fluid samples; each experiment generates a complete kinetic~curve.

The column filter consists of a glass cylinder (internal diameter 0.044~m, length 0.25~m) packed with sand at the top and bottom, and, for the column experiments considered in \S\ref{Column}, a homogeneous mixture of MRC and TMRC with an approximate mass ratio of 40:1 MRC:TMRC, in between (Figure~\ref{Batch diagram}, right). We allow for small variations in this ratio, with the TMRC fraction taken to lie in the range $(0.95/41,\ 1.05/41)$. The mixture is uniformly packed to a height of $0.1 \pm 0.005$~m, and fluid flows through the column under gravity at various low flow rates. The inlet pH is fixed at 7, and the ambient temperature is $300 \pm 3.0$~K. We collect samples at the outlet at regular intervals, measuring the  fluoride concentration, $c_\mathrm{F}^\mathrm{out}$, in these sample using an ion-selective electrode.}

\subsection{Batch modelling}

    In \citet{auton2023mathematical}, we presented a chemically based model  for MRC and TMRC, {merely asserting dominant chemical reactions. 
    Here, we carefully derive these models, appealing to the different underlying chemical mechanisms at play and  considering the many different chemical reactions that occur during the fluoride sorption process. We then appeal to chemistry to justify the choice of dominant reactions.}  
     As in \cite{auton2023mathematical}, we keep the  mechanics of the system simple, instead focusing on the underlying chemistry. We model the filter as a homogeneous porous material with material properties determined via experiments and assume the flow to be unidirectional or `plug~flow'.

Batch experiments involve a closed system, as such the only change in  $c_\mathrm{F}$  is via adsorption on the surface of the adsorbent.   Note that in batch experiments the beaker is placed in a shaker which  corresponds to a modelling assumption that the system is reaction-limited rather than transport-limited---    that is, the only change in concentration is due to the sorption.

\subsubsection{MRC: model derivation (CB-MRC model)  }
\label{MRC batch}
 \begin{figure}[htb]
    \centering  
    \includegraphics[width=1\textwidth]{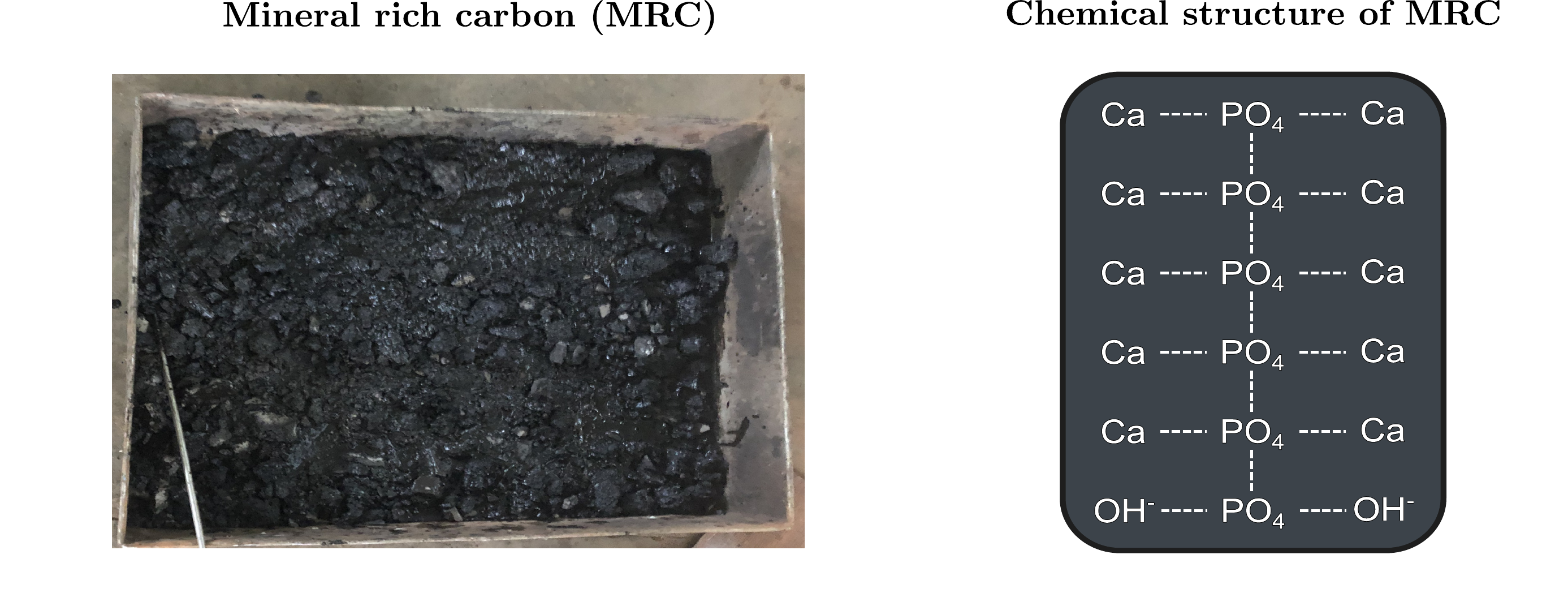}
    \caption{\small %\red{Updated 6/5/25}
    \textit{\textbf{Left:}} photo of MRC before it is sieved to have grain sizes ranging from 0.4--0.6 mm. \textit{\textbf{Right:}}  schematic of the chemical structure of MRC. 
    }
    \label{MRC_chem_schem}
\end{figure}

 As discussed earlier, \citet{medellin2014adsorption,huyen2023bone,alkurdi2019bone}, find that approximately  70\% -- 90\% of bone char (MRC) comprises hydroxyapatite and that it is the hydroxyapatite in the bone char which is predominantly responsible for the uptake of fluoride. Hydroxyapatite typically consists of  ten calcium (Ca$^{2+}$) ions,  six phosphate (PO$_4^{3-}$) molecules and two hydroxide ($\text{OH}^-$)  molecules (see Figure~\ref{MRC_chem_schem}). 

 To obtain information about hydroxyapatite such as the structure of the lattice, potential substitutions/defects (\textit{e.g.}, F$^-$ replacing OH$^-$), hydrogen bonding, water content, and ion mobility, a solid-state Nuclear Magnetic Resonance (NMR) can be conducted. \citet{mosiman2021internalization} present NMR experiments conducted on hydroxyapatite, which reveal a primary peak attributed to complexation or ion-exchange and additionally, a second spectral peak consistent with hydrogen bonding.    
Such experimental evidence contributes to the understanding of how fluoride interacts with hydroxyapatite, supporting various adsorption mechanisms proposed in the literature. The most generally accepted mechanisms are:

\begin{itemize}
    \item[i.] 
    \textbf{Complexation\footnote{Complexation is the reversible formation of a coordination complex, which is a chemical compound composed of a central atom or ion—typically a metal—that binds surrounding molecules or ions, called ligands (or complexing agents), such as fluoride, via coordinate covalent bonds.} and/or ion-exchange with OH$^-$ ions (chemisorption)}: 
\citet{mosiman2021internalization,mosiman2021probing}, and \citet{sternitzke2012uptake} conduct characterization experiments on hydroxyapatite to determine its structure, properties, and composition. 
 These experiments, including the NMR study conducted by  \citet{mosiman2021internalization},  show that OH$^-$ ions in hydroxyapatite can be replaced by F$^-$ ions, 
either through the formation of Ca--F complexes in which fluoride is coordinated by multiple Ca$^{2+}$ ions, or via a thermodynamically favourable ion-exchange mechanism. 
  \citet{sternitzke2012uptake} report that this replacement primarily occurs at the surface, while  \citet{mosiman2021internalization} argue that there is limited support for this in the literature and present evidence that both mechanisms can occur within the hydroxyapatite lattice. The ion-exchange mechanism has also been reported in numerous other works \cite[\textit{e.g.,}][]{tomar2015enhanced,balasooriya2022applications,sundaram2008defluoridation,alkurdi2019bone}.
   
    \item[ii.] \textbf{Adsorption via intermolecular forces (physisorption)}: \citet{huyen2023bone} and \citet{alkurdi2019bone} indicate electrostatic attraction between the adsorbent surface and fluoride as one possible adsorption mechanism. This mechanism can be observed under acidic conditions;  the adsorbent surface becomes positively charged via the reaction
\begin{align}
    \text{POH} + \text{H}_3\text{O}^+ &\rightleftharpoons \text{POH}_2^+ + \text{H}_2\text{O}, \label{protonation}
\end{align}
where POH represents a surface hydroxyl group on the hydroxyapatite lattice \cite{medellin2014adsorption}. However, under alkaline conditions, the surface becomes negatively charged according to the reaction
\begin{align}
    \text{POH} + \text{OH}^- &\rightleftharpoons \text{PO}^- + \text{H}_2\text{O},  \label{deprotonation}
\end{align}
in which case, \citet{medellin2014adsorption} and \citet{huyen2023bone} report that electrostatic interaction with fluoride is unlikely.

The NMR experiment conducted by \citet{mosiman2021internalization} supports the works by  \citet{sundaram2008defluoridation} and \citet{balasooriya2022applications} which  indicate that fluoride can also form hydrogen bonds with surface hydroxyl groups on hydroxyapatite.  Further, these NMR experiments also corroborate the formation of water molecules within the hydroxyapatite lattice under alkaline conditions, as described by Equation~\eqref{deprotonation}.

     \item[iii.] \textbf{Dissolution-Precipitation}:  
Numerous studies report that hydroxyapatite can dissolve, releasing calcium (Ca$^{2+}$) and phosphate (PO$_4^{3-}$) ions into the aqueous solution \cite{sternitzke2012uptake,mosiman2021internalization,mosiman2021probing,huang2023enhanced,ren2019molecular,tung2001interfacial,alkurdi2019bone}. If the solution becomes saturated with these ions and fluoride (F$^-$) is present in sufficient concentration, calcium fluoride (CaF$_2$) can precipitate, resulting in a decrease in F$^-$ concentration in the solution. However, this phenomenon has typically only been observed under strongly acidic conditions, and significantly higher fluoride concentrations are required for CaF$_2$ precipitation to occur at higher pH levels. For example, \citet{ren2019molecular} and \citet{tung2001interfacial} indicate that at pH 4, a minimum fluoride concentration of 0.005~mol/l is necessary for precipitation. Characterization studies by \citet{ren2019molecular} further show that at neutral pH, a fluoride concentration of at least 0.1~mol/l  (1900~mg/l) is needed for this precipitation to occur. This increases to 0.5~mol/l (9500~mg/l) at pH 10.
\end{itemize}

Appealing to the reactions proposed by \citet{sundaram2008defluoridation,balasooriya2022applications,tomar2015enhanced,tung2001interfacial,alkurdi2019bone}, we conclude that the ion-exchange reactions occurring within the MRC lattice can be represented as:
\begin{subequations}
\begin{align}
    \text{POH} + \text{F}^- &\rightleftharpoons \text{PF} + \text{OH}^-, \label{MRC chem 1} \\
    \text{POH}_2^+ + \text{F}^- &\rightleftharpoons \text{PF} + \text{H}_2\text{O},  \label{MRC chem 2}
\end{align}
\end{subequations}
where `P' denotes a reactive site in the MRC lattice, typically associated with phosphate groups. As reported by \citet{mosiman2021internalization}, both ion-exchange and complexation mechanisms involve substitution of OH$^-$ in the lattice and result in chemisorbed fluoride. Hence, Equations \eqref{MRC chem 1} and \eqref{MRC chem 2} can be viewed as global representations of both processes.

According to expressions proposed by \citet{sundaram2008defluoridation,balasooriya2022applications,tomar2015enhanced,medellin2014adsorption}, physical interactions (physisorption) relevant to MRC include:
\begin{subequations}
\begin{align}
    \text{POH} + \text{F}^- &\rightleftharpoons \text{POH} \cdots \text{F}^-, \label{MRC phys PO43 1} \\
    \text{POH}_2^+ + \text{F}^- &\rightleftharpoons \text{POH}_2^+ \ \text{- - -}  \   \text{F}^-, \label{MRC phys PO43 2}
\end{align}
\end{subequations}
where $\cdots$ represents hydrogen bonding and - - -     represents an electrostatic interaction.
Note that these physisorption mechanisms occur at different locations than the chemisorption processes in Equations \eqref{MRC chem 1} and \eqref{MRC chem 2}. As reported by \citet{mosiman2021internalization}, hydrogen-bonded fluoride may occupy surface or defect sites, while chemisorbed fluoride replaces lattice-bound OH$^-$ groups. The electrostatic attraction described in Equation \eqref{MRC phys PO43 2} typically occurs at the positively charged adsorbent surface under acidic conditions \cite{huang2023enhanced,alkurdi2019bone,medellin2014adsorption}.

We now turn to the evidence presented in the literature (\textit{cf.} points i, ii, and iii), as well as the operating conditions of the experimental tests conducted in this work (Section~\ref{MatandMeth}), to identify the dominant mechanisms governing fluoride removal. All experiments--- batch isotherm, batch kinetic, and column--- were initiated at pH 7. The point of zero charge (pH$_\text{ZPC}$) for MRC and TMRC are 7.6 and 6.7, respectively~\cite{chatterjee2018defluoridation}. Although the starting pH is slightly below the pH$_\text{ZPC}$ for MRC, the adsorbent surface can be considered nearly neutral at the outset. However, due to the rapid release of OH$^-$ ions, the surface quickly acquires a negative charge via deprotonation (Eq.~\ref{deprotonation}), resulting in an alkaline environment.
An immediate consequence of this shift is that the dominant fluoride species in solution remains F$^-$ over the entire operational pH range (pH~$\geq$~7)~\cite{cattarin2009electrochemical}. Further, dissolution--precipitation reactions (point iii) can be excluded as contributing mechanisms, since the maximum fluoride concentration used in this study (0.05~mol/l) remains below the threshold required for such reactions to occur at pH~$\geq$~7 (0.1~mol/l)~\cite{ren2019molecular}.

The formation of POH$_2^+$ is improbable, as it occurs primarily under acidic conditions~\cite{sarkar2006use, vithanage2012modeling}. Therefore, we neglect Equations~\eqref{MRC chem 2} and~\eqref{MRC phys PO43 2}, retaining only Equations~\eqref{MRC chem 1} and~\eqref{MRC phys PO43 1}. This simplification is supported by Figure~\ref{all_quants_outlet}, which shows an initial spike in OH$^-$ concentration that later subsides, indicating the early onset of deprotonation.

The inhibitory effect of OH$^-$ on fluoride adsorption is reflected in the reversible chemisorption reaction (Eq.~\ref{MRC chem 1}), where OH$^-$ appears on the product side. This reaction is expected to dominate over physisorption under alkaline conditions due to its greater thermodynamic and kinetic favourability.
Thus, we take the  following two reactions as the dominate reactions during the adsorption process:
\begin{subequations}
    \begin{align}
        \text{POH} + \text{F}^- &\xrightleftharpoons[k_1^\mathrm{d}]{k_1^\mathrm{a}} \text{PF} + \text{OH}^-, \label{MRC reactions Chemical} \\
        \text{POH} + \text{F}^- &\xrightleftharpoons[\kappa_2^\mathrm{d}]{k_2^\mathrm{a}} \text{POH} \cdots \text{F}, \label{MRC reactions Physical}
    \end{align}
    \label{MRC reactions}
\end{subequations}
where, $k_1^\mathrm{a}$ and $k_1^\mathrm{d}$ are the forward and reverse rate constants for the chemisorption reaction (Eq.~\ref{MRC reactions Chemical}), while $k_2^\mathrm{a}$ and $\kappa_2^\mathrm{d}$ correspond to the forward and reverse rate constants for the physisorption process (Eq.~\ref{MRC reactions Physical}). Note that $\kappa$ and $k$ are used to distinguish between rate constants with different units.

Finally, the release of OH$^-$ exclusively from the chemisorption pathway aligns with observations reported by Mosiman et al.~\cite{mosiman2021internalization}, who found that only about half of the F$^-$ uptake resulted in a net OH$^-$ loss. This is indicative of  the coexistence of both chemisorption and physisorption mechanisms.

Note that, we have neglected the explicit competitive effects between OH$^-$ and F$^-$ for adsorption sites, particularly those described by Equation~\eqref{MRC chem 2}. This assumption results from the observation that the formation of negatively charged surface species (\textit{e.g.,} PO$^-$) introduces electrostatic repulsion that reduces fluoride adsorption. Rather than modelling this repulsive interaction directly, we consolidate its effects into the overall desorption term associated with the chemisorption reaction (Eq.~\ref{MRC chem 1}), specifically through the rate constant $k_{d_1}$. The term $-k_d q_1 c_{\text{OH}}$ effectively captures the reduced fluoride adsorption capacity resulting from OH$^-$ competition and surface charge effects. By doing so, we eliminate the need to explicitly track the competitive pathway (Eq.~\ref{MRC chem 2}), which is less favourable under alkaline conditions, while still preserving the net effect on fluoride removal kinetics. 

Equation~\eqref{MRC reactions Chemical} represents a chemical reaction involving complexation/ion-exchange between F$^-$ and OH$^-$, while Equation~\eqref{MRC reactions Physical} describes a physisorption process. It is important to note that although the OH$^-$ ions interacting with F$^-$ appear identical in both reactions (\textit{i.e.,} POH is used to denote both), the adsorption mechanisms involve distinct types of sites. In Equation~\eqref{MRC reactions Chemical}, F$^-$ replaces OH$^-$ through ion-exchange at specific chemical sites. In contrast, in Equation~\eqref{MRC reactions Physical}, F$^-$ physically adsorbs onto OH$^-$ groups by occupying vacant or defect sites in the lattice.

\subsubsection{MRC: comparison with batch experiments (CB-MRC model)}
\label{MRC batch comp}

We  invoke the law of mass-action to express the reactions~(Eqs.~\ref{MRC reactions}) as a system of differential equations 
\begin{subequations}
\label{MRC_batch_eqns}
    \begin{align}
    &\frac{\partial c_\mathrm{F}}{\partial t} = -\frac{\rho^\mathrm{B}}{\phi}\left(\frac{\partial q_1}{\partial t}+\frac{\partial q_2}{\partial t}\right) \equiv -\frac{\rho^\mathrm{B}}{\phi} \frac{\partial q_\mathrm{M}}{\partial t}, \label{MRC eq system cF} \\
        &\frac{\partial c_\mathrm{OH}}{\partial t} = \frac{\rho^\mathrm{B}}{\phi}\frac{\partial q_1}{\partial t}, \label{MRC eq system cOH} \\
        &\frac{\partial q_1}{\partial t} = k_1^\mathrm{a}c_\mathrm{F}(q_1^\mathrm{m}-q_1)-k_1^\mathrm{d}c_\mathrm{OH}q_1, \label{MRC eq system q1} \\
        &\frac{\partial q_2}{\partial t} = k_2^\mathrm{a}c_\mathrm{F}(q_2^\mathrm{m}-q_2)-\kappa_2^\mathrm{d}q_2, \label{MRC eq system q2} 
    \end{align}
    \label{MRC eq system}
 \hspace{-1.5mm}where $t$ is time,   $\rho^\mathrm{B}$ is the bulk density,  $\phi$ is porosity, $c_\text{OH}$ is the concentration of hydroxide (mol$/$l), $q_1$ is the  
 moles of the adsorbate PF formed per mass of MRC  (mol/g),
$q_2$ is 
the moles of the adsorbate POH$\cdots$F formed per mass of MRC (mol/g),
$q_i^\mathrm{m}$ is the maximum attainable value of $q_i$, for $i=1,2$ with $q_\mathrm{M}\defeq q_1+q_2$. 
The  bulk density of an adsorbent $\rho^\mathrm{B}$ is defined to be the total initial mass of the adsorbent, divided by the total volume of the filter, and the porosity $\phi$ is defined to be the fluid fraction of the batch filter. Thus, for batch filters  the quantity $\gamma \defeq \rho^\mathrm{B} /\phi$  is equivalent to the dose. As such, for the isotherm experiments $\gamma\defeq\gamma_\text{iso}\equiv 7$ and for the kinetic experiments $\gamma\defeq\gamma_\text{kin}\equiv 1$. Note that the same doses are used regardless of adsorbent and thus, for batch experiments, the value of $\gamma$ depends only on isotherm \textit{vs} kinetic and not the particular adsorbent~used.

Initially, the MRC is contaminant free; we add a solution to the beaker which contains a known, initial, concentration of F$^-$ ($c_\mathrm{F}^\mathrm{i}$) and an initial concentration of OH$^-$ ($c_\mathrm{OH}^\mathrm{i}$) which is calculated from the pH  of the system. Thus, at
 $t=0$:
\begin{align}
    q_1=q_2 &=0\\
    c_\mathrm{F} &=c_\mathrm{F}^\mathrm{i}\\
    c_\mathrm{OH} &=c_\mathrm{OH}^\mathrm{i}
\end{align}
\end{subequations}
where $c_\mathrm{\star}^\mathrm{i}$ is the inlet concentration of contaminant  where $\star$ denotes either F or OH. 
We refer to the 
chemically based  model for MRC as defined in Equations~(\ref{MRC eq system}) as the CB-MRC model.
%\editmarker{} 

\begin{figure}[htb!]
    \centering
    \includegraphics[width=.95\textwidth]{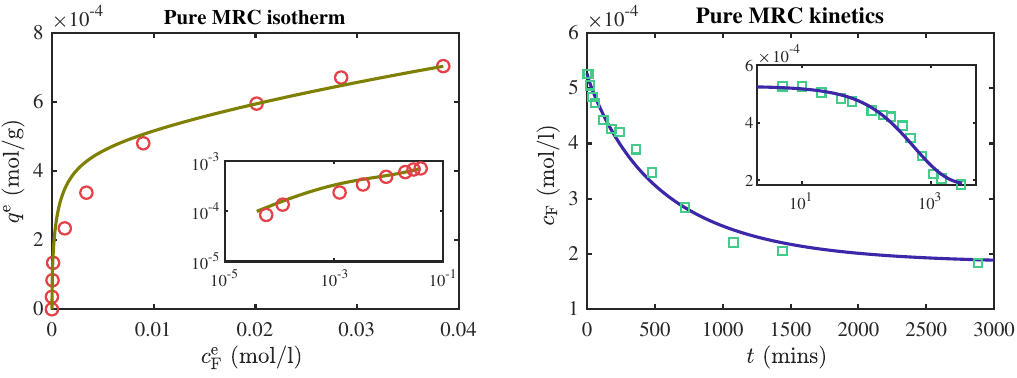}
     \caption{\small %\red{Updated 5/5/25}
     Isotherm (left) and  kinetic study (right) for MRC fitted with the  CB-MRC model  (Eqs.~\ref{MRC eq system})
     \textit{\textbf{Left}}: All experimental data points (red circles) are the equilibrium values of $q$ against $c_\mathrm{F}$ for individual experiments, 
     each with  a different, known $c_\mathrm{F}^\mathrm{i}$. 
      To examine the fit more closely for small $c_\text{F}^\text{e}$, we also include a log-log plot of $q^\text{e}$ against $c_\text{F}^\text{e}$  (inset). 
     \textit{\textbf{Right}}:~All experimental data points (green squares) are from the same experiment. 
     To examine the fit more closely at small times, we include a semi-log plot of $c_\text{F}$ against $t$ (inset). 
    \label{MRC global opt batch}
 }
\end{figure}

For the CB-MRC model we solve for the isotherm analytically but the kinetic function numerically in \verb|MATLAB|$^\text{\copyright}$ using \verb|ode15s|. 
The isotherm is the equilibrium of the system~(Eqs.~\ref{MRC_batch_eqns}) given by $q_\mathrm{M}=q_\mathrm{M}^\mathrm{e}\equiv q_1^\mathrm{e}+q_2^\mathrm{e}$. 
At equilibrium Equations~(\ref{MRC_batch_eqns}) reduce to
\begin{subequations}
    \begin{align}
        k_1^\mathrm{a}(c_\mathrm{F}^\mathrm{i}-\gamma_\mathrm{iso} q_\mathrm{M}^\mathrm{e})(q_1^\mathrm{m}-q_1)&=k_1^\mathrm{d}(c_\mathrm{OH}^i+\gamma_\mathrm{iso}  q_1^\mathrm{e})q_1\\
        k_2^\mathrm{a}(c_\mathrm{F}^\mathrm{i}-\gamma_\mathrm{iso}  q_\mathrm{M}^\mathrm{e})(q_2^\mathrm{m}-q_2)&=\kappa_2^\mathrm{d}q_2
    \end{align}
\end{subequations}
with solution
\begin{subequations}
\begin{align}
    q_1^\mathrm{e} &\defeq \frac{-\left(c_\mathrm{OH}^\mathrm{i}+K_1c_\mathrm{F}^\mathrm{e}\right)+\sqrt{\left(c_\mathrm{OH}^\mathrm{i}+K_1c_\mathrm{F}^\mathrm{e}\right)^2+4\gamma_\mathrm{iso} K_1c_\mathrm{F}^\mathrm{e}q_1^\mathrm{m}}}{2\gamma_\mathrm{iso}}\\ 
    q_2^\mathrm{e} &\defeq \frac{\mathcal{K}_2c_\mathrm{F}^\mathrm{e}q_2^\mathrm{M}}{1+\mathcal{K}_2c_\mathrm{F}^\mathrm{e}}
\end{align}
where 
\begin{equation}
        K_1\defeq \frac{k_1^\mathrm{a}}{k_1^\mathrm{d}}\qquad \text{and} \qquad                 \mathcal{K}_2\defeq \frac{k_2^\mathrm{a}}{\kappa_2^\mathrm{d}}
\end{equation}
\end{subequations}
Further,  we  use the constraint,
\begin{equation}
    q_\mathrm{M}^\mathrm{e}= \frac{c_\mathrm{F}^\mathrm{i}-c_\mathrm{F}^\mathrm{e}}{\gamma_\mathrm{kin}},
\end{equation}
 obtained via integrating  Equation~(\ref{MRC eq system cF}), to find $\mathcal{K}_2$ using the kinetic data: 
\begin{equation}
\label{K2}
    \mathcal{K}_2 \equiv \frac{c_\mathrm{F}^\mathrm{i}-c_\mathrm{F}^\mathrm{e}-\gamma_\mathrm{kin}q_1^\mathrm{e}}{c_\mathrm{F}^\mathrm{e}\left[\gamma_\mathrm{kin}\left(q_2^\mathrm{m}+q_1^\mathrm{e}\right)+c_\mathrm{F}^\mathrm{e}-c_\mathrm{F}^\mathrm{i}\right]}. 
\end{equation}

Figure~\ref{MRC global opt batch} presents the experimental data alongside the best fit as obtained from the model defined by Equations~(\ref{MRC_batch_eqns}). The left panel of Figure~\ref{MRC global opt batch} corresponds to the MRC isotherm, while the right panel shows the results of the MRC kinetic study.
 
The MRC isotherm  requires  three fitting parameters: $K_1$, $q_\mathrm{M}^\mathrm{m}$  and $q_2^\mathrm{m}/q_\mathrm{M}^\mathrm{m}$. These three parameters are fitted for first using just the isotherm. The kinetic curve then requires two further fitting parameter: $k^\mathrm{a}_1$ and $k^\mathrm{a}_2$.
Note that,  the three  parameters which can be determined using only the isotherm are 
intrinsic constants of the system and are subsequently used in the column experiments reducing the required number of fitting parameters  for the column filter. 
The remaining two MRC parameters, $k_1^\mathrm{a}$ and $k_2^\mathrm{a}$, are only used for the kinetic curves. These two parameters must be re-fit for in the column experiments as they are specific to the experimental setup. 
Our fitting is conducted in \verb|MATLAB|$^\text{\copyright}$ using the \verb|GlobalSearch| function to minimise the objective function  which is the absolute difference between the model predictions and data points, for the isotherm and then subsequently the kinetic curve. 
Note that \verb|MATLAB|$^\text{\copyright}$'s \verb|GlobalSearch| algorithm, which is based on \citet{ugray2007scatter},
 utilises a stratified-sampling procedure to randomly generate trial points \cite{locatelli1999random}.
 As a result, each time the same code is run, slightly different optima 
 are produced. To produce the values in Tables~\ref{MRC_table} and \ref{TMRC_table} we have run the optimisation algorithm multiple times ($\mathcal{O}$(10)) and 
 selected the parameter values which yield the lowest SSE and highest~R$^2$.

\begin{table}[htb]
    \centering \hspace{-8mm}\footnotesize{
    \begin{tabular}{|c|c|c||c|c|c||c|c|c|}
        \hline 
        \multicolumn{9}{|c|}{\textbf{MRC isotherm and kinetic parameters }} \\ 
        \hline
        \multicolumn{3}{|c||}{\textbf{Extracted (kinetic)}} & \multicolumn{3}{|c||}{\textbf{ Optimised parameters}} & \multicolumn{3}{|c|}{\textbf{Goodness of fit }} \\
        \hline
        Param. & Value & Units & Param. & Value & Units & Param. & Iso./Kin. & Value \\
        \hline  \hline
        $c_\mathrm{F}^\mathrm{i}$ & 5.26$\times 10^{-4}$ \ \ (10) & mol/l  \ (mg/l) & $ K_1$& 4.74 & -- & SSE & Isotherm & 0.0514  \\
        \hline
        $c_\mathrm{OH}^\mathrm{i}$ & $1 \times10^{-7}$ \ \ (0.0017) & mol/l \ (mg/l)& $q_\mathrm{M}^\mathrm{m}$ & 0.00174 & mol/g & R$^2$ & Isotherm & 0.961\\ 
        \hline
        $c_\mathrm{F}^\mathrm{e}$ & 1.84$\times 10^{-4}$ \ \ (3.5) & mol/l\  (mg/l) & $ q_2^\mathrm{m}/q_\mathrm{M}^\mathrm{m}$  & 0.729 & -- & SSE & Kinetic & 0.0105  \\
        \hline \cline{1-3}
        \multicolumn{3}{|c||}{\textbf{Calculated}} &$k_1^\mathrm{a} $ &$2.78$  & l/(mol·s) & R$^2$  & Kinetic   & 0.986 \\
        \hline \cline{1-3} %\cline{7-9}
        $\mathcal{K}_2$ & 6.00 & l/mol& $k_2^\mathrm{a}$ & 0.389& l/(mol·s) &  \multicolumn{3}{|c|}{} \\
       \hline% \cline{7-9}
    \end{tabular}
   }
   \caption{%\red{UPDATE 26/6/25} 
   \label{MRC_table} %\luc{Have updated the values 2025}
    \small{Parameters for the chemically based MRC model (CB-MRC, Eqs.~\ref{MRC eq system}), where $q_\mathrm{M}^\mathrm{m}\defeq q_1^\mathrm{m}+q_2^\mathrm{m}$. \textit{\textbf{Left (top)}}: parameter values extracted from the kinetic experiment; $c_{\mathrm{OH}}^\mathrm{i}$ is calculated based on the pH of the system which is held at 7. The concentrations are shown in mol/l and in brackets the equivalent in mg/l. \textit{\textbf{Left (bottom):}}  the value of $\mathcal{K}_2$ calculated using the optimised parameters and Equation~(\ref{K2}). \textit{\textbf{Centre:}} Parameters optimised using both the isotherm and kinetic experiments. \textit{\textbf{Right:}} %The goodness of fit parameters for  the model: 
    the sum of squares error (SSE) and the coefficient of determination (R$^2$). 
    The R$^2$ is nondimenisonal by construction.  For the kinetic study we nondimensionalised the SSE with $c_\mathrm{F}^\mathrm{i}$, while for the isotherm, we nondimensionalised the SSE with the largest experimentally determined value of $q^e$.  Note that to convert $c_\mathrm{F}$ from mol/l to mg/l we must multiply by 19000, while to convert $c_\mathrm{OH}$ we must multiply by 17000.
    } 
    }
    \label{MRC new model table}
\end{table}

%%%%%%%%%%%%%%%%%%%%%%%%%

Table~\ref{MRC new model table} presents the parameter values extracted from the kinetic study (top left) and those calculated from the fitting parameters (bottom left), alongside the five optimized batch parameters (center) and the goodness-of-fit metrics (right). 
Notably, the initial hydroxide ion concentration, $c_\mathrm{OH}^\mathrm{i}$, is fixed at $10^{-7}$ mol/l, reflecting the system’s pH of 7. The pH is defined chemically as $\mathrm{pH} = -\log [\mathrm{H}_3\mathrm{O}^+]$. The hydroxide ion concentration is related to the pOH by $c_\mathrm{OH} = 10^{-\mathrm{pOH}}$, where $\mathrm{pOH} = -\log [\mathrm{OH}^-]$. 

The relationship between pH and pOH derives from the self-ionization of water,
\begin{equation}
\label{SelfIonisation}
2\mathrm{H}_2\mathrm{O} \rightleftharpoons \mathrm{H}_3\mathrm{O}^+ + \mathrm{OH}^-,
\end{equation}
which at room temperature ($\sim 25^\circ$C) satisfies the approximate relation
\begin{equation}
\label{pHofOH}
\mathrm{pH} + \mathrm{pOH} \approx 14.
\end{equation}
As the initial pH is recorded as  7, it follows that $\mathrm{pOH} = 7$, and thus the hydroxide ion concentration is $c_\mathrm{OH}^\mathrm{i} = 10^{-7}$ mol/l \cite{chang2009chemistry}.

%%%%%%%%%%%%%%%%%%%%%%%%%

 The equilibrium constant $ K_1 $ is order 1, neither particularly high nor low; being nondimensional, this clearly suggests a reversible reaction. In contrast, $ \mathcal{K}_2 $ is dimensional, which complicates its direct interpretation. However, given that $ \mathcal{K}_2 $ scales with $1/c_{\mathrm{F}}^{\mathrm{e}}$ and  
 that  $ c_{\mathrm{F}}^{\mathrm{e}} $ is very low, one would expect $ \mathcal{K}_2 $ to be quite large. The fact that it is instead of order 10 suggests that $ q_2^{\mathrm{e}} $ is also small. 
 This observation is consistent with the presence of a negatively charged surface under alkaline conditions, which tends to hinder physisorption due to its relatively weak interaction forces~\cite{medellin2014adsorption,huang2023enhanced}.
 Further, the ratio $q_2^{\mathrm{m}} / q_M^{\mathrm{m}} = 0.637 $ indicates a slightly greater number of vacant sites in the hydroxyapatite lattice available for fluoride physisorption compared to hydroxide ions available for ion-exchange. This ratio being close to one half aligns with the findings of \citet{mosiman2021internalization}, who reported that the net loss of OH$^-$ is approximately half of the F$^-$ uptake. This would imply that physisorption accounts for roughly half of the maximum fluoride removal capacity, with ion-exchange accounting for the remainder.

Regarding model fitting, the coefficient of determination R$^2 $ is, by definition, nondimensional for both the isotherm and kinetic curves. The sum of squared errors (SSE), however, carries units of mol$^2$/g$^2$ for the isotherm and mol$^2$/l$^2$ for the kinetic study. To facilitate comparison, the SSE was nondimensionalized: for the kinetic model, this was achieved using the initial concentration $ c_\mathrm{F}^{\mathrm{i}} $, and for the isotherm, using the largest data point in the dataset. These adjustments allow for a meaningful assessment of model performance. The combination of an R$^2 $ value close to 1 and relatively low nondimensional SSE indicates a good fit overall as we have here.  However, despite the strong values for SSE and R$^2$, in the isotherm some deviation is observed around $ c_\mathrm{F}^{\mathrm{e}} \approx 0.005 $, where the curvature of the model slightly diverges from the experimental data (see Figure~\ref{MRC global opt batch}).
We hypothesise that this discrepancy may stem from the accumulation of small, systematic errors---     for example those related to dosage precision and experimental uncertainties in measurement and mixing. These seemingly minor inaccuracies, when summed, can lead to noticeable deviations in model performance at specific concentration ranges. Further investigation into error propagation and sensitivity analysis could help quantify the extent to which these small errors influence model accuracy.

\subsubsection{TMRC: model derivation (IE-TMRC model)} 
\label{TMRC batch}
 \begin{figure}[htb]
    \centering  
    \includegraphics[width=1\textwidth]{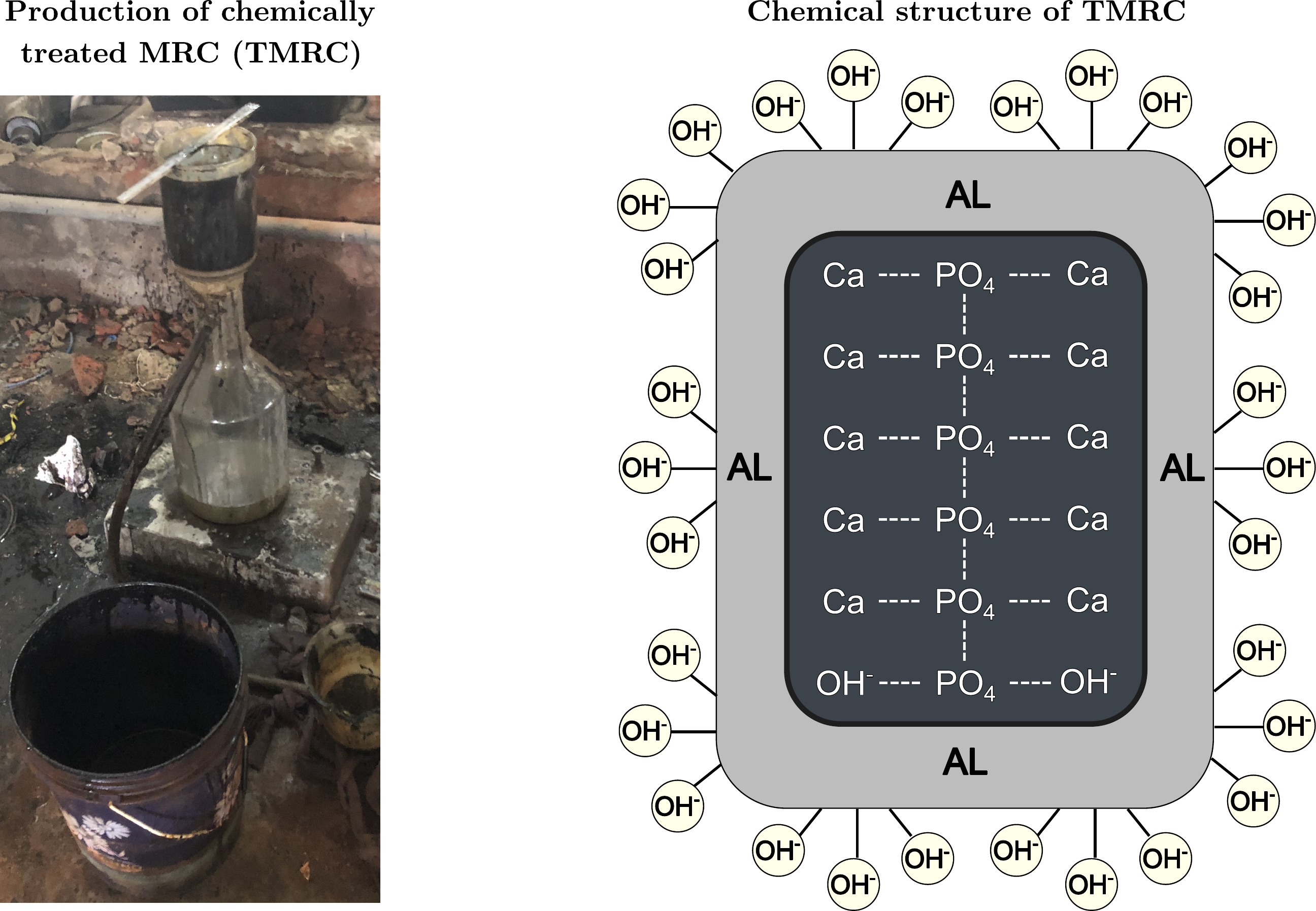}
    \caption{\small %\red{upadted 6/5/25}
    Left: photo showing the prototype procedure for making TMRC from MRC. Right: schematic of the chemical structure of TMRC; showing how the  fluoride and hydroxide molecules interact with the TMRC. 
}
    \label{TMRC_chem_schem}
\end{figure}

Chemically-treated mineral rich carbon (TMRC) is made by grinding down MRC and coating it in  aluminium hydroxide $\big(\text{Al(OH)}_3\big)$. This significantly aids in the adsorption of fluoride; the adsorption capacity of TMRC is approximately ten times greater than that of MRC \cite{chatterjee2018defluoridation}. Thus, we  neglect any reaction that occurs in TMRC and does not involve aluminium, including the contribution of the hydroxyapatite core. 

The aluminium hydroxide coating layer consists of a crystalline structure that can form coordination complexes 
with water molecules~\cite{hem1967form,nordin1999mechanisms,craig2017assessing,lin2020role}. The primary chemical and physical mechanisms affecting fluoride adsorption onto the Al(OH)\textsubscript{3} surface are largely analogous to those proposed for hydroxyapatite. These mechanisms include:

\begin{itemize}
    \item[i.] \textbf{Complexation and/or ion-exchange with OH$^-$ (chemisorption)}: Several studies have reported that fluoride interacts with aluminium hydroxide-based adsorbents, including activated alumina~\cite{craig2017assessing,lin2020role,hao1986adsorption} and various aluminium hydroxide polymorphs~\cite{nordin1999mechanisms,gong2012effect}, primarily through surface complexation. This process can alter the local crystalline structure. However, Gong et al.~\cite{gong2012effect} and Ghorai et al.~\cite{ghorai2004investigations} observed that this mechanism predominately occurs at low pH. Under alkaline conditions, fluoride adsorption occurs mainly via ion-exchange between F$^-$ and surface OH$^-$ groups, a process that typically preserves the structural integrity of the coating. This ion-exchange mechanism has also been described by~\cite{gong2012adsorption,jia2015fluoride}. When the OH$^-$ groups involved are part of a pre-existing surface complex, as is the case here,   this is sometimes referred to as `ligand exchange'~\cite{tripathy2006removal,gai2022defluoridation,chen2022removal,barathi2014graphene}.
    
    \item[ii.] \textbf{Adsorption via intermolecular forces (physisorption)}:  Similar to the mechanism described for MRC, 
        the surface of the TMRC  can undergo protonation in acidic media and deprotonation in alkaline conditions, via analogous reactions to those involving POH  but instead involving Al--OH (see reactions~(Eq.~\ref{protonation}) and (Eq.~\ref{deprotonation})). 
    Under acidic conditions, electrostatic attraction may facilitate fluoride physisorption. However, under alkaline conditions, such interactions are significantly weakened due to surface deprotonation and electrostatic repulsion~\cite{ghorai2004investigations,lin2020role,biswas2007adsorption,chen2022removal,craig2017assessing,gong2012adsorption,nordin1999mechanisms}. Additionally, hydrogen bonding between fluoride and surface hydroxyl groups has been proposed as a contributing physisorption mechanism~\cite{nordin1999mechanisms,lin2020role}.

    \item[iii.] \textbf{Dissolution–precipitation}: It is widely accepted that aluminium-fluoride complexes (\textit{e.g.,} AlF$_\alpha^{3-\alpha}$) are more soluble under acidic conditions, with minimal dissolution occurring near neutral pH~\cite{nordin1999mechanisms,craig2017assessing,chen2022removal,lin2020role,gong2012effect}. However, studies using activated alumina~\cite{lin2020role,craig2015comparing} and various aluminium hydroxide phases~\cite{benezeth2008dissolution,yang2007surface} have reported increasing dissolution of fluorohydroxoaluminate species (\textit{e.g.}, AlF$_\alpha$(OH)$_\beta^{3-\alpha-\beta}$) with increasing pH. Among these, Al(OH)$_4^-$ is often the dominant species. This process is described by the equilibrium reaction:
    \begin{align}
        \text{Al(OH)}_{3\mathrm{(cr)}} + \text{OH}^- &\rightleftharpoons \text{Al(OH)}_4^-, \label{AlOH4dissolution}
    \end{align}
    where Al(OH)\textsubscript{3(cr)} represents the crystalline solid phase of the coating, while OH$^-$ and Al(OH)$_4^-$ are aqueous species in the surrounding fluid. This dissolution equilibrium is influenced by the competition between OH$^-$ and F$^-$ ions for surface adsorption sites~\cite{chen2022removal}, and may account for the elevated aluminium concentrations observed in the column effluents, as reported by Chatterjee et al.~\cite{chatterjee2018defluoridation}.
\end{itemize}

Analogous to the model development for MRC, we exclude all reactions associated with a positively charged surface. This is justified by the initial pH of 7 and pH$_{\text{ZPC}}$ = 6.7 for TMRC~\cite{chatterjee2018defluoridation}, which suggests that the adsorbent surface becomes negatively charged under operating conditions. Further, the anticipated release of OH$^-$ ions is expected to rapidly drive the system into a strongly alkaline environment. For this reason, we assume that the predominant fluoride species in solution remains as F$^-$ over the entire operating pH range (pH $\geq$ 7)~\cite{cattarin2009electrochemical}.

In line with assumptions made in deriving MRC model, we assume chemisorption occurs via an ion-exchange or ligand-exchange mechanism between OH$^-$ and F$^-$ at the aluminium hydroxide surface of the TMRC. Appealing to \citet{ghorai2004investigations,tripathy2006removal,gai2022defluoridation,chatterjee2018defluoridation}, this reaction can be represented as:
\begin{equation}
    \text{Al--OH} + \text{F}^- \xrightleftharpoons[k_\mathrm{T}^\mathrm{d}]{k_\mathrm{T}^\mathrm{a}} \text{Al--F} + \text{OH}^-, \label{TMRC reactions}
\end{equation}
where $k_\mathrm{T}^\mathrm{a}$ and $k_\mathrm{T}^\mathrm{d}$ are the forward and reverse rate constants, respectively. Here, Al--OH represents individual surface bonds between aluminium and hydroxide groups that are displaced by Al--F bonds during fluoride adsorption. Although ion-exchange and ligand-exchange differ mechanistically, both can be described using this global reaction, which defines the chemisorption mechanism outlined in point (i).

Physisorption via hydrogen bonding (point ii) was considered in the MRC model, based on the assumption that the hydroxyapatite lattice contains vacancies and defects capable of hosting F$^-$ through hydrogen bonding with OH$^-$ groups~\cite{mosiman2021internalization}. However, TMRC consists of a thin crystalline coating layer, where hydrogen bonding would be limited to surface interactions. Given that electrostatic attraction is already suppressed under alkaline conditions due to surface deprotonation, we assume that surface-level hydrogen bonding is insufficiently strong to contribute meaningfully to fluoride adsorption. As such, physisorption is neglected in the TMRC model.

Finally, for consistency with the MRC model and to simplify the analysis, we neglect competition between F$^-$ and OH$^-$ for surface adsorption sites. This implies that dissolution processes occurring at high pH---   described in point (iii) and governed by reaction~(Eq.~\ref{AlOH4dissolution})---   are not included in the model.

% %%%%%%%%%%%%%%%%%%%%%%%%%%%

%%%%%%%%%%%%%%%%%%%%%%%%%%%%%%%%%%%%%
\subsubsection{TMRC: comparison with batch experiments (IE-TMRC model)}
\label{TMRC batch comp}
We again invoke the law of mass-action to express reaction~(Eq.~\ref{TMRC reactions}) as a system of differential equations~\cite{nie2012enhanced,hao1986adsorption},
\begin{subequations}
\label{TMRC_batch_equations}
    \begin{align}
           &\frac{\partial c_\mathrm{F}}{\partial t} = -\frac{\rho^\mathrm{B}}{\phi}\frac{\partial q_\mathrm{T}}{\partial t}, \label{TMRC eq system cF}\\
        &\frac{\partial c_\mathrm{OH}}{\partial t} = \frac{\rho^\mathrm{B}}{\phi}\frac{\partial q_\mathrm{T}}{\partial t},\label{TMRC eq system cOH} \\ 
        &\frac{\partial q_\mathrm{T}}{\partial t} = k_\mathrm{T}^\mathrm{a}c_\mathrm{F}(q_\mathrm{T}^\mathrm{m}-q_\mathrm{T})-k_\mathrm{T}^\mathrm{d}c_\mathrm{OH}q_\mathrm{T}, 
        \label{ODE}
    \end{align}
    \label{TMRC eq system}

\noindent where $q_\mathrm{T}$ is the number of moles of the adsorbate $\text{Al--F}$ formed per mass of TMRC (mol/g), and 
$q_\mathrm{T}^\mathrm{m}$ is the maximum attainable value of $q_\mathrm{T}$. As  for MRC   
 $\gamma \defeq \rho^\mathrm{B} /\phi$ with $\gamma\defeq\gamma_\text{iso}\equiv 7$ for the isotherm experiments  and $\gamma\defeq\gamma_\text{kin}\equiv 1$ for the kinetic experiments and, as before,  at $t=0$,
    \begin{align}
q_\mathrm{T}&=0, \\ c_\mathrm{F}&=c_\mathrm{F}^\mathrm{i}\\ 
c_\mathrm{OH}&=c_\mathrm{OH}^\mathrm{i}. 
    \end{align}
\end{subequations}

As in the case of MRC, the inhibitory effect of OH$^-$ on fluoride adsorption is inherently accounted for by the presence of OH$^-$ on the product side of the reversible chemisorption reaction~(Eq.~\ref{TMRC reactions}). However, for TMRC, it is possible that some of the released hydroxide is not present as free OH$^-$ ions, but rather as hydroxyl groups complexed in the form of Al(OH)$_4^-$, as suggested by Equation~\eqref{AlOH4dissolution}. This interpretation is consistent with the findings reported by Chatterjee et al.~\cite{chatterjee2018defluoridation}.

\begin{figure}[tb]
    \centering
    \includegraphics[width=.95\textwidth]{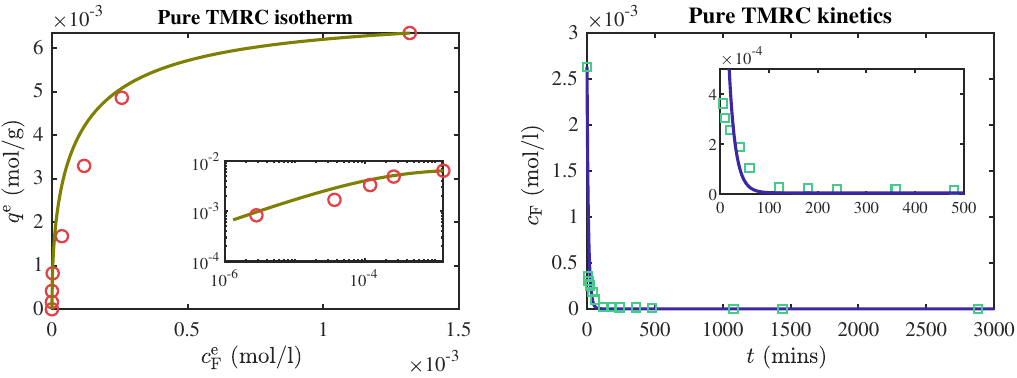}
    \caption{\small %\red{Updated 5/5/25}
    Isotherm (left) and  kinetic study (right) for TMRC fitted with  both the  IE-TMRC model (Eqs.~\ref{TMRC eq system}).  
    \textit{\textbf{Left:}} To examine the fit more closely for small $c_\text{F}^\text{e}$, we also include a log-log plot of $q^\text{e}$ against $c_\text{F}^\text{e}$  (inset). 
     \textit{\textbf{Right}}:
     To examine the fit more closely at small times, we include a semi-log plot of $c_\text{F}$ against $t$ (inset).      }
    \label{TMRC global opt batch}
\end{figure}
  
    \begin{table}[b]
    \centering
    \footnotesize{
    \begin{tabular}{|c|c|c||c|c|c||c|c|c|}
        \hline 
        \multicolumn{9}{|c|}{\textbf{TMRC isotherm and kinetic parameters }} \\ 
        \hline
        \multicolumn{3}{|c||}{\textbf{Extracted (kinetic)}}  & \multicolumn{3}{|c||}{\textbf{Optimised parameters}}& \multicolumn{3}{|c|}{\textbf{Goodness of fit}}\\
        \hline
        Param. & Value & Units  & Param. & Value & Units& Param. & Iso./Kin. & Value \\
        \hline \hline
        $c_\mathrm{F}^\mathrm{i}$ & 2.63$\times 10^{-3}$ \ \ (50.0) & mol/l \ (mg/l)& 
       % \cellcolor[rgb]{0.7922,0.8471, 0.8706}
        $q_\mathrm{T}^\mathrm{m}$  & 
        %\cellcolor[rgb]{0.7922,0.8471, 0.8706}
        0.0069 & 
       % \cellcolor[rgb]{0.7922,0.8471, 0.8706}
        mol/g &
        %\cellcolor[rgb]{0.7922,0.8471, 0.8706}
        SSE & 
        %\cellcolor[rgb]{0.7922,0.8471, 0.8706}
        Isotherm & 
        %\cellcolor[rgb]{0.7922,0.8471, 0.8706}
        0.0615 \\
        \hline
        $c_\mathrm{OH}^\mathrm{i}$ & $1 \times10^{-7}$ \ \ (0.0017) & mol/l \ (mg/l)& 
        %\cellcolor[rgb]{0.7922,0.8471, 0.8706} 
         $k_\mathrm{T}^\mathrm{a}$ & 
       % \cellcolor[rgb]{0.7922,0.8471, 0.8706}
        $16.5$& 
       % \cellcolor[rgb]{0.7922,0.8471, 0.8706}
         l/(mol·s)& 
        %\cellcolor[rgb]{0.7922,0.8471, 0.8706} 
        R$^2$ & 
       % \cellcolor[rgb]{0.7922,0.8471, 0.8706}
        Isotherm& 
        %\cellcolor[rgb]{0.7922,0.8471, 0.8706}
        0.938 \\ 
        \hline  \cline{4-6} 
        $c_\mathrm{F}^\mathrm{e}$ & 4.21$\times 10^{-6}$ \ \ (0.08) & mol/l \ (mg/l)& 
        %\cellcolor[rgb]{1,1, 0.8}
      \multicolumn{3}{|c||}
      {\textbf{Calculated}}  & 
      %  \cellcolor[rgb]{1,1, 0.8} 
        SSE & 
        %\cellcolor[rgb]{1,1, 0.8}
        Kinetic& 
       % \cellcolor[rgb]{1,1, 0.8}
       0.283  \\
        \hline \cline{4-6}
        \multicolumn{3}{|c||}{} & $K_\mathrm{T}$ & 384 & -- & R$^2$ & Kinetic &0.680 \\
        \hline  
    \end{tabular}}
    \caption{%\red{UPDATED 26/6/25}
    \label{TMRC_table}\small 
    Parameters of the ion-exchange model for TMRC (IE-TMRC, Eqs.~\ref{TMRC eq system}). The same groups of parameters are presented as in Table~\ref{MRC_table} (\textit{i.e.,} Extracted, Optimised parameters, Calculated and Goodness of fit). Further, 
    the goodness of fit parameters have been nondimensionalised as in Table~\ref{MRC_table} 
    }
    \label{TMRC new model table}
\end{table}

We refer to the ion-exchange  model for TMRC as defined in Equations (\ref{TMRC eq system}) as the IE-TMRC model. 
For the IE-TMRC model, we solve for both the isotherms and kinetic functions  analytically. For the kinetic result we express  
Equation~(\ref{ODE}) as 
\begin{subequations}
\label{TMRC_kinetic_ana}
\begin{equation}
  \frac{\mathrm{d}q_\mathrm{T}}{\mathrm{d}t} = A\left(q_\mathrm{T}-R_+\right)\left(q_\mathrm{T}-R_-\right)  
\end{equation}
where 
\begin{equation}
    A\defeq \gamma_\mathrm{kin}(k_\mathrm{T}^\mathrm{a}-k_\mathrm{T}^\mathrm{d}) \qquad \text{and} \qquad
    R_\pm \defeq \frac{B\pm\sqrt{B^2-4AD}}{2A}
\end{equation}
with 
\begin{equation}
    B\defeq k_\mathrm{T}^\mathrm{a}(\gamma_\mathrm{kin} q_\mathrm{T}^\mathrm{m}+c_\mathrm{F}^\mathrm{i})+k_\mathrm{T}^\mathrm{d}c _\mathrm{OH}^\mathrm{i}\qquad \text{and} \qquad
    D\defeq k_\mathrm{T}^\mathrm{a}c_\mathrm{F}^\mathrm{i}q_\mathrm{T}^\mathrm{m}. 
\end{equation}
Solving Equations~(\ref{TMRC_kinetic_ana}) yields 
\begin{equation}
\label{TMRCqT}
    q_\mathrm{T} = \frac{R_-R_+\left\{\exp{\left[-A(R_+-R_-)t\right]}-1\right\}}{R_-\exp{\left[-A(R_+-R_-)t\right]}-R_+}, 
\end{equation}
where  $c_\mathrm{F}$ is given by 
\begin{equation}
c_\mathrm{F}= c_\mathrm{F}^\mathrm{i}-\gamma_\mathrm{T}q_\mathrm{kin}. 
\end{equation}
\end{subequations}
The isotherm is determined via the equilibrium of Equation~(\ref{ODE}), yielding
\begin{subequations}
\begin{equation}
        q^\mathrm{e}_\mathrm{T} = -f_1(c_\mathrm{F}^\mathrm{e})+\sqrt{\left[f_1(c_\mathrm{F}^\mathrm{e})\right]^2+f_2(c_\mathrm{F}^\mathrm{e})}
    \end{equation}
     where 
   % \luc{Define $K_\mathrm{T}^a$ and $\gamma$ and qeT}
    \begin{equation}
    f_1(c_\mathrm{F}^\mathrm{e}) \defeq \frac{c_{\mathrm{OH}}^\mathrm{i}+K_\mathrm{T}c_\mathrm{F}^\mathrm{e}}{2\gamma_\mathrm{iso}}, \qquad 
    f_2(c_\mathrm{F}^\mathrm{e}) \defeq \frac{q_\mathrm{T}^\mathrm{m}K_\mathrm{T}c_\mathrm{F}^\mathrm{e}}{\gamma_\mathrm{iso}}\qquad \text{and} \qquad
      K_\mathrm{T} \defeq \frac{k_\mathrm{T}^\mathrm{a}}{k_\mathrm{T}^\mathrm{d}}
    \end{equation}
\end{subequations}
We use the constraint \begin{equation}
\gamma_\mathrm{kin}q_\mathrm{T}^\mathrm{e}=c_\mathrm{F}^\mathrm{i}-c_\mathrm{F}^\mathrm{e}
\end{equation}
which results from integrating Equation~(\ref{TMRC eq system cF})
to determine $K_\mathrm{T}$ in terms of the other parameter 
\begin{equation}
  \label{KT} K_\mathrm{T}=\frac{\left(c_\mathrm{F}^\mathrm{i}-c_\mathrm{F}^\mathrm{e}\right)\left(c_\mathrm{F}^\mathrm{i}-c_\mathrm{F}^\mathrm{e}+c_\mathrm{OH}^\mathrm{i}\right)}{c_\mathrm{F}^\mathrm{e}\left(\gamma_\mathrm{kin}q_\mathrm{T}^\mathrm{m}-c_\mathrm{F}^\mathrm{i}+c_\mathrm{F}^\mathrm{e}\right)}.
\end{equation}

Figure~\ref{TMRC global opt batch} shows the experimental data and the best fit as predicted by the model defined in Equations~(\ref{TMRC eq system}). The left panel of Figure~\ref{TMRC global opt batch} corresponds to the TMRC isotherm, while the right panel shows the results of the TMRC kinetic study.

There are two fitting parameters for TMRC: $q_\mathrm{T}^\mathrm{m}$ which is fitted from the isotherm, and $k_\mathrm{T}^\mathrm{a}$ which is subsequently fitting from the kinetic study. As before,  all fittings are via the \verb|GlobalSearch| function in \verb|MATLAB|$^\text{\copyright}$ such that the objective function  is the absolute difference between the model predictions and data points. 
As for the CB-MRC model, $q_\mathrm{T}^\mathrm{m}$ is an intrinsic constant of the the system, while $k_\mathrm{T}^\mathrm{a}$ must be re-fit for the column experiments as it is specific to the experimental set-up.

Table~\ref{TMRC_table} presents the parameter values extracted from the kinetic study (left) and those calculated from the fitting parameters (centre, bottom), alongside the two optimized batch parameters (centre, top) and the goodness-of-fit metrics (right). 

Given that $K_{\mathrm{T}}$ is dimensionless and relatively large (order of hundreds), it suggests that the chemisorption reaction undergone by the  TMRC is relatively direct, with a stronger forward (adsorption) tendency than reverse (desorption), implying a quasi-irreversible behaviour. This interpretation is consistent with experimental observations indicating that TMRC exhibits a greater affinity for fluoride compared to MRC~\cite{chatterjee2018defluoridation}, effectively functioning as a more efficient adsorbent. Further, the maximum adsorption capacity $q_{\mathrm{T}}^{\mathrm{m}}$ is approximately one order of magnitude greater than $q_{\mathrm{M}}^{\mathrm{m}}$, which aligns with the reported finding that TMRC absorbs roughly ten times more  than MRC~\cite{chatterjee2018defluoridation}.

Figure~\ref{TMRC global opt batch} (left panel) demonstrates a good fit of the equilibrium isotherm to the experimental data, as indicated by a low sum of squared errors (SSE = 0.0615) and a high coefficient of determination (R$^2$ = 0.938). The right panel shows the kinetic profile, which also aligns reasonably well with the observed data. However, the early-time region exhibits a rapid transition, indicating that the majority of fluoride uptake occurs within a short initial period. This sharp adsorption front introduces potential measurement uncertainties, as the rate of fluoride removal may exceed the temporal resolution of the sampling process. We hypothesise that this initial transient is responsible for the comparatively higher SSE (0.283) and the moderate R$^2$ value (0.68) observed 
in the kinetic fit. Nevertheless, it is important to note that the parameters obtained from these
batch experiments yield a good agreement when applied to the column experiments, supporting the robustness of the model.

\section{Column filter}

\label{Column}

%\vspace{-1cm}

We consider the steady flow of  fluid contaminated with F$-$ through a column filter, modelled as a homogeneous porous medium. We assume unidirectional (plug) flow, as such the  fluid flow moves parallel to
the $z$-axis (aligned vertically downward) with a constant Darcy velocity $v$. The contaminated water enters uniformly at the inlet, $z = 0$, and exits at the outlet, $z = L$. The inlet concentrations of fluoride and hydroxide ions 
are controlled, while the outlet fluoride concentration, $c_\mathrm{F}^\mathrm{out}$, is measured as a function of time; the graph of $c_\mathrm{F}^\mathrm{out}$ \textit{vs} $t$ 
  \begin{wrapfigure}{r}{0.27\textwidth}
  \centering
  \includegraphics[width=0.25\textwidth]{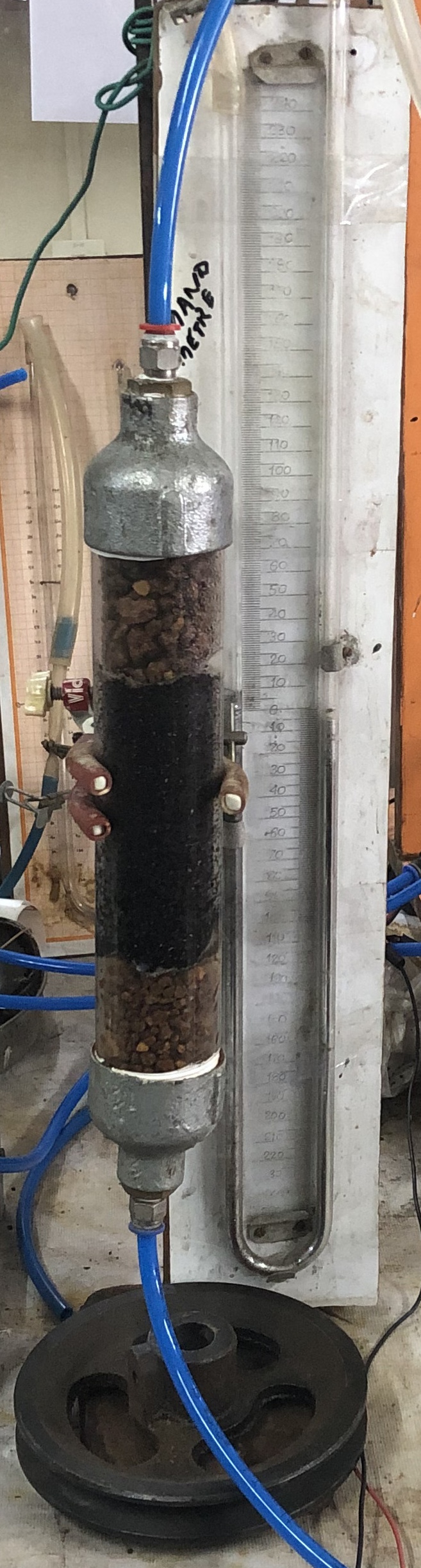}
  \caption{\small Column lab experimental setup. }
    \vspace{-4cm}
\end{wrapfigure}
is known as a breakthrough curve. Note that, in this section we consider concentrations solely in units of mg/l rather than mol/l; the conversion between the units is the same as detailed in Table~\ref{MRC_table}.

  In column filters, the dominant mechanisms for contaminant transport are advection and hydrodynamic (shear) dispersion. For the MRC–TMRC mixture, we assume 
  fluoride removal occurs via the same chemical reactions as in the batch system, namely those described by Equations~(\ref{MRC reactions}) and~(\ref{TMRC reactions}) \cite{ghorai2004investigations,barathi2014graphene,tomar2015enhanced,sundaram2008defluoridation,alkurdi2019bone}. Accordingly, the  sink terms remain identical to those used in the batch model.

Transport of F$^-$ and OH$^-$ ions through the filter is modelled by a system of advection--diffusion--reaction equations with removal via both MRC and TMRC components:
\begin{subequations}
   %\label{column}
    \begin{align}
    \label{Combo_cf}
    &\frac{\partial c_\mathrm{F}}{\partial t} = D_\mathrm{F}\frac{\partial^2c_\mathrm{F}}{\partial z^2}-v\frac{\partial c_\mathrm{F}}{\partial z} - \left(\frac{\rho^\mathrm{B}_\mathrm{M}}{\phi_\mathrm{C}}\frac{\partial q_1}{\partial t}+\frac{\rho^\mathrm{B}_\mathrm{M}}{\phi_\mathrm{C}}\frac{\partial q_2}{\partial t}+\frac{\rho^\mathrm{B}_\mathrm{T}}{\phi_\mathrm{C}}\frac{\partial q_\mathrm{T}}{\partial t}\right),\\
         \label{Combo_cOH}&\frac{\partial c_\mathrm{OH}}{\partial t} = D_\mathrm{OH}\frac{\partial^2c_\mathrm{OH}}{\partial z^2}-v\frac{\partial c_\mathrm{OH}}{\partial z} + \left(\frac{\rho^\mathrm{B}_\mathrm{M}}{\phi_\mathrm{C}}\frac{\partial q_1}{\partial t}+\frac{\rho^\mathrm{B}_\mathrm{T}}{\phi_\mathrm{C}}\frac{\partial q_\mathrm{T}}{\partial t}\right), \\ 
        \label{Combo_q1} &\frac{\partial q_1}{\partial t} = k_1^\mathrm{a}c_\mathrm{F}(q_1^\mathrm{m}-q_1)-k_1^\mathrm{d}c_\mathrm{OH}q_1, \\
         \label{Combo_q2}&\frac{\partial q_2}{\partial t} = k_2^\mathrm{a}c_\mathrm{F}(q_2^\mathrm{m}-q_2)-\kappa_2^\mathrm{d}q_2, \\
         \label{Combo_qT}&\frac{\partial q_\mathrm{T}}{\partial t} = k_\mathrm{T}^\mathrm{a}c_\mathrm{F}(q_\mathrm{T}^\mathrm{m}-q_\mathrm{T})-k_\mathrm{T}^\mathrm{d}c_\mathrm{OH}q_\mathrm{T}, 
    \end{align}
    \label{Column eq system}

\noindent where $D_\mathrm{F}$ and $D_\mathrm{OH}$ denote the effective diffusivities of F$^-$ and OH$^-$, respectively, and $\phi_\mathrm{C}$ is the porosity of the MRC--TMRC mixture, given by $\phi_\mathrm{C} \defeq (1-f)\phi_\mathrm{M} + f\phi_\mathrm{T}$, with $f$ representing the TMRC fraction in the mixture. The effective diffusivities are not the static molecular diffusivity which would be $\mathcal{O}(10^{-9})$, rather they  incorporate the effect of slow flow dispersion, hence `effective'. The numeric values for these are taken from  \citet{levenspiel1998chemical}, to be $D_\mathrm{F} = D_\mathrm{OH} = 2.9 \times 10^{-7}$ m$^2$/s.

The filter is assumed to be initially free of F$^-$ ions, but not of OH$^-$ ions. Thus, at $t = 0$, the initial conditions are: 
\begin{equation}
    q_1 = 0, \quad q_2 = 0, \quad q_\mathrm{T} = 0, \quad c_\mathrm{F} = 0, \quad \text{and} \quad c_\mathrm{OH} = c_\mathrm{OH}^\mathrm{f}.
\end{equation}

\noindent Following the analysis in \citet{pearson1959note} and as discussed in \citet{aguareles2023mathematical}, we impose Dankwert's boundary condition at the inlet and a zero flux condition at the outlet: %the following boundary conditions:
\begin{equation}
\label{BC_dim}
    {v}c_\ddagger(0,t)-D_\ddagger\left.\frac{\partial c_\ddagger}{\partial z}\right|_{z=0}={v}c_{in}, \quad \text{and } \quad
\left.\frac{\partial c_\ddagger}{\partial z}\right|_{z=L}=0, \quad \text{for all } t, 
\end{equation}
\end{subequations}

\noindent where $\ddagger$ denotes either F or OH,  and $c_\text{in}$ represents the corresponding inlet concentration of F$^-$.  We refer to the model for the MRC–TMRC mixture, defined by Equations~(\ref{Column eq system}), as the CB-MT model. All numerical solutions are obtained using \verb|MATLAB|$^\text{\copyright}$’s built-in initial-boundary value problem solver, \verb|pdepe|.

\begin{figure}[tb]
    \centering
 \includegraphics[width=.95\textwidth]{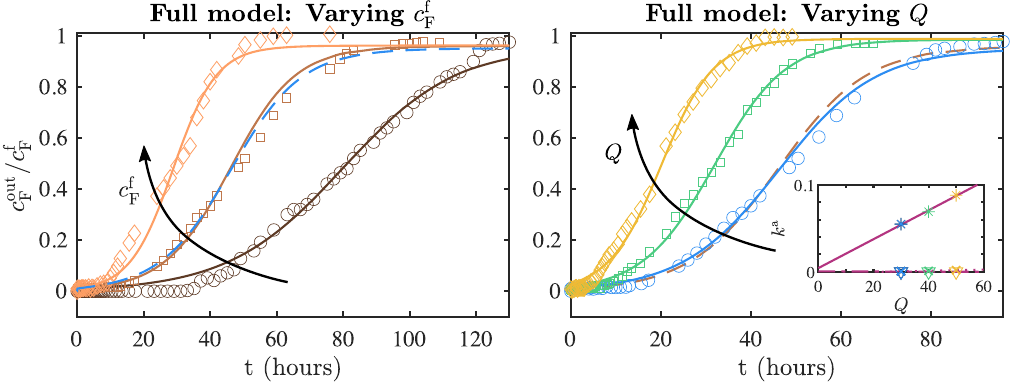}
    \caption{\small %\red{UPDATED 9/5/25} 
Breakthrough curves for a column filter composed of a mixture of MRC and TMRC in an approximate mass ratio of 40:1 (MRC:TMRC), compared to the CB-MT model predictions (\textit{cf.} Equations~\ref{Column eq system}). \textbf{Left:} Experimental data and model fits for three inlet fluoride concentrations, $c_\mathrm{F}^\mathrm{f} = 5.44$\,mg/l (circles), 9.5\,mg/l (squares), and 14.5\,mg/l (diamonds), at a fixed flow rate of 30\,l/day. \textbf{Right:} Experimental and model results for three flow rates, $Q\in\{30$\,l/day,  40\,l/day  50\,l/day$\}$ (circles, squares and diamonds, respectively),  with an approximate $c_\mathrm{F}^\mathrm{f}=10$ mg$/$l (see Table~\ref{TMRC MRC model table} for precise values). The inset on the right panel displays the fitted values of the forward reaction rates: $k_\mathrm{T}^\mathrm{a}$ (asterisks), $k_1^\mathrm{a}$ (downward facing triangles), and $k_2^\mathrm{a}$ (crosses) and a linear best fit for each $k^\mathrm{a}$.  To highlight the repeated experiment,  dashed lines indicate the model fit based on the other data set: the brown curve from the left panel is shown as a dashed brown line in the right panel, and the blue curve from the right is shown as dashed blue in the left.
    \label{combo 3 concentration breaktrhough}}
\end{figure}

To assess the performance of the CB-MT model, we consider two sets of column breakthrough data: (i) three breakthrough curves with varying inlet fluoride concentrations $c_\mathrm{F}^\mathrm{f} \in \{5 \pm 0.5\ \text{mg}/\text{l}, 10 \pm 0.5\ \text{mg}/\text{l}, 15 \pm 0.5\ \text{mg}/\text{l}\}$ at a fixed flow rate of 30 l/day (Figure~\ref{combo 3 concentration breaktrhough}, left, brown to orange), and (ii) three breakthrough curves with varying flow rates $Q \in \{30\ \text{l/day}, 40\ \text{l/day}, 50\ \text{l/day}\}$ at a fixed fluoride concentration of 10$\pm$0.5 mg/l (Figure~\ref{combo 3 concentration breaktrhough}, right, blue to yellow). One experiment is repeated across both sets (highlighted in grey in Tables~\ref{TMRC MRC model table} and \ref{TMRC model table}). In all experiments, the mass ratio of MRC to TMRC is held approximately constant at 40:1, and the filter height is around 0.1~m.

As discussed in the last paragraph of \S\ref{MRC batch}, the model is sensitive to small variations in input parameters. To account for this, we permit slight deviations in certain experimental conditions during the fitting process. Specifically, the TMRC fraction, $f$, which is nominally 1/41, is allowed to vary within the range $f \in (0.95/41, 1.05/41)$, while the column height $L$ such that $L \in (0.095, 0.105)$m. Inlet fluoride concentrations are also fitted within their experimental uncertainty of $\pm$0.5mg/l. All other parameters are held fixed.

   \hspace{-1cm}\begin{table}[htb]
 \centering 
    \footnotesize{
    \begin{tabular}{|c||c|c|c||c|c|c||c|}
        \hline 
        \multicolumn{8}{|c|}{\textbf{Optimised and goodness of fit  parameters for column filter: Full model }} \\ 
        \hline    \multirow{ 2}{*}{Param.}& \multicolumn{3}{|c||}{\textbf{ Feed concentration (approx)}}& \multicolumn{3}{|c||}{\textbf{Flow rate}}& \multirow{ 2}{*}{Units}\\
       \cline{2-7}     &  5 mg/l & \cellcolor[rgb]{0.7922,0.8471, 0.8706} 10 mg/l &  15 mg/l & \cellcolor[rgb]{0.7922,0.8471, 0.8706} 30 l/day & 40 l/day &  50 l/day &  \\
 \hline \hline
        $k_\mathrm{T}^\mathrm{a}$   & \multicolumn{3}{|c||}{0.0594} & 0.0548    &  0.0691    &  0.0880 &l/(mol·s) \\
          \hline
  $k_1^\mathrm{a}$    & \multicolumn{3}{|c||}
  {2.19$\times10^{-4}$} & 2.73$\times10^{-4}$   &  1.01$\times10^{-4}$  & 
  1.11$\times10^{-4}$  & l/(mol·s) \\
 \hline
 $k_2^\mathrm{a}$  & \multicolumn{3}{|c||}{$2.03\times10^{-4}$}& $5.29\times10^{-4}$		& $1.02\times10^{-3}$   &  $1.89\times10^{-3}$  & l/(mol·s) \\     \hline  \hline
 \multirow{ 2}{*}{$c_\mathrm{F}^\mathrm{f}$} & 5.44  &   9.50  &  14.5  & 9.50  &   9.84   &  10.5  &  mg/l \\
\cline{2-8}  & $2.86\times10^{-4}$  &   $5.00\times10^{-4}$  &  $7.63\times10^{-4}$  & $5.00\times10^{-4}$  &  $5.18\times10^{-4}$   &  $5.53\times10^{-4}$  & mol/l \\
 \hline
 $f$ &  1.05/41  &   1.05 /41 &  1.04/41 &  1.05/41  &   1.02 /41 &  0.950/41  & -- \\  \hline
  $L$ &  0.105  &   0.105  &  0.101 & 
  %Lbopt = 1.049999999999977   1.032170235382149   0.950000000002836
  0.105  &   0.104  &  0.0950  & m \\  \hline
   \hline
 SSE & 0.06316   &   0.03098    &    0.04832 &  0.02234  &    0.01490   & 0.02548 & -- \\
 \hline
    R$^2$   &  	0.9930	& 0.9956	& 0.9916 &  	0.9966	& 0.9978	& 0.9962
 & -- \\
\hline 
    \end{tabular}
    \caption{\small %\red{UPDATED VALUES 12/7/25}
Fitted parameters and corresponding goodness-of-fit metrics for the chemically based combined MRC  and TMRC model (CB-MT, Eqs.~\ref{Combo_cf}). The fraction of TMRC is denoted $f$ and the bed height $L$. Note that the SSE has been normalised using the precise inlet concentration (\textit{i.e.,} the values of $c_\mathrm{F}^\mathrm{f}$ given in this table).}
    \label{TMRC MRC model table}}
\end{table}

Figure~\ref{combo 3 concentration breaktrhough} compares the model predictions with experimental breakthrough data; the left panel shows results for varying $c_\mathrm{F}^\mathrm{f}$ and the right panel for varying flow rate $Q$. The CB-MT model uses the intrinsic  parameters  as determined from  the isotherm fitting: $K_1$, $q_\mathrm{M}^\mathrm{m}$, $q_2^\mathrm{m}/q_\mathrm{M}^\mathrm{m}$, $q_\mathrm{T}^\mathrm{m}$. Their values are given in Tables~\ref{MRC new model table} and \ref{TMRC new model table}. The three kinetic adsorption parameters $k_\mathrm{T}^\mathrm{a}$, $k_1^\mathrm{a}$, and $k_2^\mathrm{a}$ are treated as fitting parameters for the column study. For each individual experiment, we also fit $f$, $L$, and $c_\mathrm{F}^\mathrm{f}$ within their respective bounds. 

Fitting is performed by minimising the SSE using \verb|MATLAB|$^\text{\copyright}$’s \verb|lsqcurvefit| function, which employs local optimisation. Due to the high dimensionality of the parameter space, this approach is significantly more computationally efficient than global optimisation methods. To ensure convergence to physically reasonable values, parameter ranges were manually constrained before optimisation. 

For the set of experiments with varying $c_\mathrm{F}^\mathrm{f}$, the forward reaction rates, $k^\mathrm{a}$ are fitted globally across all three breakthrough curves, due to the assumption that the $k^\mathrm{a}$ values are independent of inlet fluoride concentration. In contrast, for the set of experiments with varying $Q$, the $k^\mathrm{a}$ values are fitted individually since it is widely reported that the forward reaction rates depend on flow rate \cite{myers2023development,wakao1978effect,patmonoaji2023dissolution,inglezakis2020liquid,misic1982liquid};
The inset on the right panel of Figure~\ref{combo 3 concentration breaktrhough} shows the fitted values for the three $k^\mathrm{a}$ with a linear best fit.

As shown in Figure~\ref{combo 3 concentration breaktrhough}, the CB-MT model accurately captures the breakthrough behaviour across all six experiments. The fit is particularly notable in the varying-concentration scenario, where a single set of kinetic parameters suffices to describe all three curves.  Dashed lines highlight the repeated experiment across the two data sets, underscoring the model's consistency and predictive capability.

\begin{figure}[b]
    \centering
 \includegraphics[width=1\textwidth]{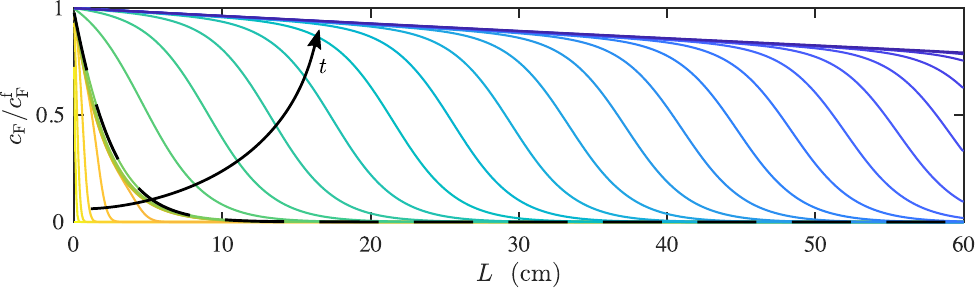}
    \caption{\small %\red{updated 29/5/25} 
    Normalised fluoride concentration is shown as a function of position along the filter for increasing times (yellow to blue).
    Here, we consider a filter of 0.6~m instead of the $\sim$0.1~m filter used for the experiments; a longer computational domain is considered to capture the full evolution of the concentration  profile. Parameter values correspond to the case of varying feed concentration, with an inlet fluoride concentration of approximately 10~mg/l. Specifically, we use: $k_\mathrm{T}^\mathrm{a} = 0.0594$, $k_1^\mathrm{a} = 2.19 \times 10^{-4}$, $k_2^\mathrm{a} = 2.03 \times 10^{-4}$, $c_\mathrm{F}^\mathrm{f} = 9.5$~mg/l, $f = 1.05/41$, and $L = 0.105$~m. The black dashed line marks the transition in time discretisation: time points before this line are spaced logarithmically, and those after are spaced linearly.
    }
    \label{notTW}
\end{figure}

Note that, the relative magnitudes of the forward reaction rates $k^\mathrm{a}$ differ between the column and batch studies, which is expected given the distinct flow regimes. In both setups, $k_1^\mathrm{a},\ k_2^\mathrm{a} \ll k_\mathrm{T}^\mathrm{a}$;  however in the batch experiments $k_1^\mathrm{a}$ is an order of magnitude greater than $k_2^\mathrm{a}$, while in the column experiments $k_2^\mathrm{a}$ ranges from the same order of magnitude as $k_1^\mathrm{a}$ to an order of magnitude greater than $k_1^\mathrm{a}$.
One possible explanation is that the ion-exchange mechanism active in TMRC may interfere with or suppress the ion-exchange process in MRC, thereby reducing the effective rate associated with MRC and altering the apparent values of $k_1^\mathrm{a}$ and~$k_2^\mathrm{a}$.

The quality of the model fit is high across all cases, as evidenced by R$^2$ values exceeding 0.991 for all six breakthrough curves. Additionally, the sum of squared errors (SSE) remains below 0.063 in all cases, indicating that the cumulative deviation between the experimental data and model predictions is less than 6.32\% of $c_\mathrm{F}^\mathrm{f}$. For the case of varying $c_\mathrm{F}^\mathrm{f}$---     where a global fit was performed across all three curves---     the SSE values are slightly higher as expected, ranging from 0.0310 to 0.0632. In contrast, for the case involving varying flow rate $Q$, the SSE values are smaller, ranging from 0.0149 to 0.0255. Together, the consistently high $R^2$ values and low SSE confirm the robustness and accuracy of the model fitting.

Figure~\ref{notTW} shows the spatial concentration profile of fluoride along the filter length over increasing times (from yellow to blue). Two distinct timescales are evident. The first, associated with TMRC and one MRC reaction ($q_2$), exhibits a near-travelling wave behaviour. The second, associated with the slower MRC physisorption process ($q_1$), manifests as a gradual diagonal front visible in the later-time contours (blue shades).

\begin{figure}[tb]
    \centering
 \includegraphics[width=1\textwidth]{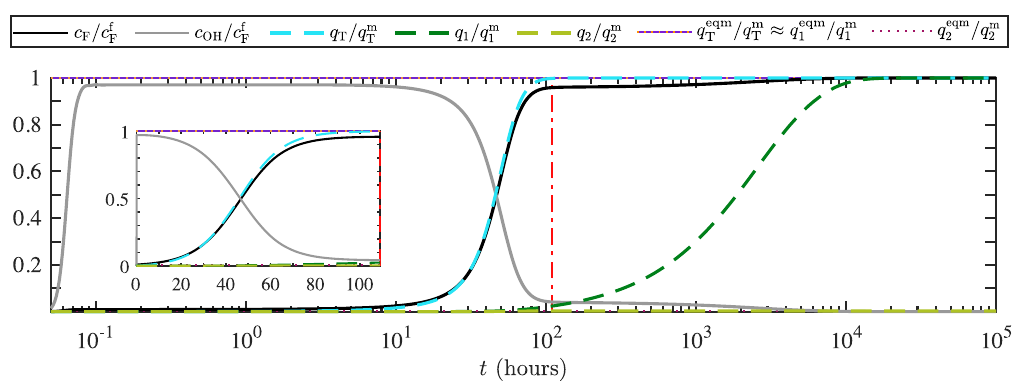}
    \caption{\small %\red{NEW: Updated 28/5/25:  }
    Outlet quantities are shown on a logarithmic time scale, using the same parameter values as in Figure~\ref{notTW}. The equilibrium values of $q_\star$, where $\star$ denotes either T, 1, or 2,   are denoted by $q_\star^\mathrm{eqm}$. The maximum value of the ratio $c_\mathrm{OH}/c_\mathrm{F}^\mathrm{f} \approx 0.97$ corresponds to a peak pH of 10.7. The inset displays the same five outlet quantities plotted on a linear time scale for comparison. The vertical red dot-dash line is fixed at 109 hours, at which point $q_2$ has reached over 95\% of its equilibrium value, and $q_\mathrm{T}$ has reached approximately 99.8\%. 
    \label{all_quants_outlet}}
\end{figure}

To further understand these dynamics, Figure~\ref{all_quants_outlet} presents the evolution of all normalised outlet quantities, $c_\mathrm{F}/c_\mathrm{F}^\mathrm{f}$, $c_\mathrm{OH}/c_\mathrm{F}^\mathrm{f}$, $q_\mathrm{T}/c_\mathrm{T}^\mathrm{m}$, $q_1/q_1^\mathrm{m}$, and $q_2/q_2^\mathrm{m}$ versus time with the fitting parameters taken to be $k_\mathrm{T}^\mathrm{a} = 0.0594$, $k_1^\mathrm{a} = 2.19 \times 10^{-4}$, $k_2^\mathrm{a} = 2.03 \times 10^{-4}$, $c_\mathrm{F}^\mathrm{f} = 9.5$~mg/l, $f = 1.05/41$, and $L = 0.105$~m, as corresponds to the 10~mg/l experiment in the set of experiments with varying $c_\mathrm{F}^\mathrm{f}$. 
Initially, the outlet concentration of OH$^-$ increases due to its production as a by-product of the chemical reactions. As $c_\mathrm{OH}$ begins to decline, a small amount of fluoride begins to appear in the outflow. Once the decline of $c_\mathrm{OH}$ accelerates, $c_\mathrm{F}^\mathrm{out}$ rises more rapidly. This marks the point at which the $q_2$ reaction becomes active; however, $q_2$ quickly saturates to its equilibrium value, which remains far below its maximum isotherm-defined capacity, confirming that the impact of $q_2$ on fluoride removal is minimal.
 Instead of $q_2$ converging to its maximum isotherm-defined capacity,  $q_2$  converges to its equilibrium value:
\begin{equation}
q_2^\mathrm{eqm} = \frac{q_2^\mathrm{m} k_2^\mathrm{a} c_\mathrm{F}^\mathrm{f}}{\kappa_2^\mathrm{d} + k_2^\mathrm{a} c_\mathrm{F}^\mathrm{f}}.
\end{equation}
The minimal role played by $q_2$ aligns with previous findings by~\citet{medellin2014adsorption} and~\citet{huyen2023bone}, who report that physisorption is strongly inhibited on negatively charged surfaces due to electrostatic repulsion.

As time progresses and the outlet concentration of OH$^-$ vanishes, the filter nears saturation. The majority of fluoride removal is attributed to TMRC ($q_\mathrm{T}$), as evidenced by the close alignment between the cyan dashed and black solid curves in Figure~\ref{all_quants_outlet}. The contribution from $q_1$ is marginal and becomes significant only at much later times, when it gradually saturates. At the point marked by the vertical red dot-dashed line (109 hours), $q_2$ and $q_\mathrm{T}$ have reached over 95\% and 99.8\% of their respective equilibrium values, while only 2.3\% of $q_1$'s maximum adsorption capacity is utilised. Full saturation of $q_1$ (to 99\%) occurs only after approximately 12,500 hours, indicating its action over a much longer timescale.
This slow filling of $q_1$ also explains the diminishing amplitude observed in the concentration profiles in Figure~\ref{notTW} at late times.

The maximum pH observed in the system is 10.7, occurring at $t \approx 3.2$ hours. This pH lies just above the threshold for enhanced solubility of alumina in the presence of fluoride. According to Craig et al.~\cite{craig2015comparing}, dissolution of aluminum hydroxide is negligible in the pH range 4--10, while Lin et al.~\cite{lin2020role} report that fluoride significantly increases alumina solubility at pH $>$ 9. Hence, the observed maximum pH supports the assumption that dissolution in TMRC can be neglected for most of the relevant operating range, validating Equation.~\eqref{AlOH4dissolution}.

Analysis of all five key outlet quantities, in combination with the reaction rate parameters ($k_1^\mathrm{a}, k_2^\mathrm{a} \ll k_\mathrm{T}^\mathrm{a}$; see Table~\ref{TMRC MRC model table}), clearly indicates that TMRC dominates fluoride removal, despite MRC being present in a much larger quantity (by mass). This is consistent with previous experimental findings by Chatterjee et al.~\cite{chatterjee2018defluoridation}, who report that TMRC exhibits more than ten times the adsorption capacity of MRC.

Given this dominance, we now consider a reduced model that includes only TMRC-based adsorption. In the following section, we fit this reduced model to experimental breakthrough curves.

\section{Reduced model}

\begin{figure}[b]
    \centering
 \includegraphics[width=.95\textwidth]{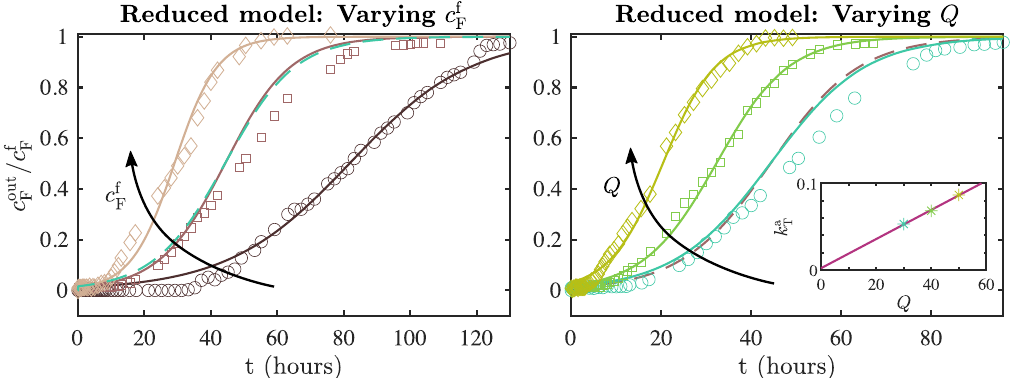}
    \caption{\small %\red{updated 8/5/25}
 Breakthrough curves fitted using the reduced model (Eqs.~\ref{Combo_cf}, \ref{Combo_cOH}, and \ref{Combo_qT}), showing the same experimental data as in  Figure~\ref{combo 3 concentration breaktrhough}.  
\textbf{Left:} Experimental data and model fits for three inlet fluoride concentrations, $c_\mathrm{F}^\mathrm{f} = 5.15$\,mg/l (circles), 9.5\,mg/l (squares), and 14.5\,mg/l (diamonds), at a fixed flow rate of 30\,l/day. 
\textbf{Right:} Experimental and model results for three different flow rates, $Q \in \{30, 40, 50\}$\,l/day (circles, squares, and diamonds, respectively), with an inlet fluoride concentration of approximately 10\,mg/l (see Table~\ref{TMRC model table}). The inset shows the fitted values of the forward reaction rate constant $k_\mathrm{T}^\mathrm{a}$ along with a linear best-fit trend. As in Figure~\ref{combo 3 concentration breaktrhough}, to emphasise the repeated experiment, model fits from one panel are shown as dashed lines in the other.
    \label{combo 3 concentration breaktrhough TMRC}}
\end{figure}

Motivated by the dynamics shown in Figure~\ref{all_quants_outlet}, we now consider a reduced model that neglects both components of the MRC. This simplification is justified by the minimal impact of $q_1$ and $q_2$ on fluoride removal, as discussed above. The resulting reduced CB-MT model comprises Equations~(\ref{Combo_cf}), (\ref{Combo_cOH}), and (\ref{Combo_qT}), where we take 
\[
\frac{\partial q_1}{\partial t} = 0, \quad \frac{\partial q_2}{\partial t} = 0.
\]

\begin{table}[htb]
    \centering
    \footnotesize{
    \begin{tabular}{|c||c|c|c||c|c|c||c|}
        \hline 
        \multicolumn{8}{|c|}{\textbf{Optimised and goodness of fit  parameters for column filter: reduced model }} \\ 
        \hline    \multirow{2}{*}{Param.}& \multicolumn{3}{|c||}{\textbf{ Feed concentration (approx)}}& \multicolumn{3}{|c||}{\textbf{Flow rate}}& \multirow{ 2}{*}{Units}\\
       \cline{2-7}     &  5 mg/l & \cellcolor[rgb]{0.7922,0.8471, 0.8706} 10 mg/l &  15 mg/l & \cellcolor[rgb]{0.7922,0.8471, 0.8706} 30 l/day & 40 l/day &  50 l/day &  \\
 \hline \hline
        $k_\mathrm{T}^\mathrm{a}$   & \multicolumn{3}{|c||}{0.0569} & 
        0.0530 		&	0.0683   &  0.0866 &l/(mol·s) \\
             \hline  \hline 
\multirow{2}{*}{$c_\mathrm{F}^\mathrm{f}$} & 5.15 &   9.50   &  14.5  & 9.50  &   9.68    &  10.5  &  mg/l \\
\cline{2-8}  & $2.71\times10^{-4}$  &   $5.00\times10^{-4}$  &  $7.63\times10^{-4}$  & $5.00\times10^{-4}$  &  $5.09\times10^{-4}$   &  $5.50\times10^{-4}$  & mol/l \\ 
 \hline 
 $f$ &  1.05/41  &   1.05 /41 &  1.05/41 &  1.05/41  &   1.04 /41 &  0.950/41  & -- \\  \hline %[  1.049999999999978   1.039790495164475   0.975945758344599]
  $L$ &  0.105  &   0.105  &  0.105 &  0.105  &   0.104  &  0.0984  & m \\
              \hline  \hline
 SSE &  0.06301  &   0.1163    &   0.05196 &  0.1032  &   0.01648    & 0.02884 & -- \\
 \hline
    R$^2$   &  	0.9930	& 0.9834	& 0.9910 &  	0.9844	& 0.9976	&0.9957
 & --  \\
\hline 
    \end{tabular}}
    \caption{%\red{updated 12/7/25}
    \small 
    Fitted parameters and corresponding goodness-of-fit metrics for the reduced model (Eqs.~\ref{Combo_cf}, \ref{Combo_cOH}, and \ref{Combo_qT}). As in Table~\ref{TMRC MRC model table}, the sum of squared errors (SSE) is normalised using the exact inlet fluoride concentration, $c_\mathrm{F}^\mathrm{f}$, as listed in this table. 
    }
    \label{TMRC model table}
\end{table}

This reduction in model complexity significantly decreases the number of parameters requiring calibration: only one fitting parameter, $k_\mathrm{T}^\mathrm{a}$ is needed for the breakthrough experiments (in addition to one from the isotherm), compared to three in the full model (in addition to four from the isotherm).    We assess the performance of this reduced model by fitting it to the same experimental data set, which spans a range of $c_\mathrm{F}^\mathrm{f}$ and $Q$. 

Figure~\ref{combo 3 concentration breaktrhough TMRC} presents the model fits obtained using the reduced model, with the corresponding fitting parameters listed in Table~\ref{TMRC model table}. Visually, the fits are slightly less accurate than those obtained with the full model but still capture the dominant fluoride removal mechanisms effectively. The inset in Figure~\ref{combo 3 concentration breaktrhough TMRC} again shows the fitted values of $k_\mathrm{T}^\mathrm{a}$ as a function of flow rate $Q$, along with a linear best fit. As before, the approximately linear trend is consistent with established correlations in the literature~\cite{myers2023development, Myer20b, wakao1978effect, patmonoaji2023dissolution, inglezakis2020liquid, misic1982liquid}.

For the reduced model, R$^2$ exceeds 0.983 for all breakthrough curves, with four of the six achieving R$^2 > 0.991$. The normalised SSE  ranges from 0.0165 to 0.119. These consistently high R$^2$ values and low SSE values confirm the robustness and reliability of the reduced model in describing the system's behaviour.

Moreover, the fitted values of the operational parameters $c_\mathrm{F}^\mathrm{f}$, $f$, and $L$ are broadly consistent between the full and reduced models. 
This consistency further indicates that the reduced model correctly captures the key removal dynamics, and supports the notion that observed discrepancies in the input parameters reflect true deviations in the experimental setup rather than modelling artefacts.

\begin{figure}[tb]
    \centering
 \includegraphics[width=1\textwidth]{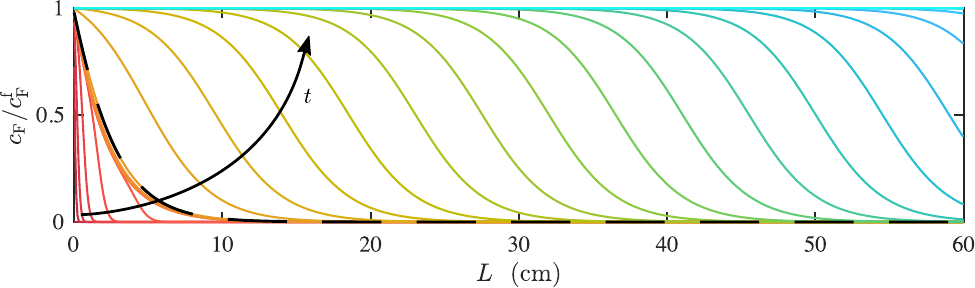}
    \caption{\small %\red{updated 2/6/25 } 
Normalised fluoride concentration as predicted by the reduced model 
as a function of position along the filter for increasing times (red to blue). We use the parameters determined for the case of varying inlet fluoride concentration around $c_\mathrm{F}^\mathrm{f} \sim 10$\,mg/l---     that is, $k_\mathrm{T}^\mathrm{a} = 0.0569$, $c_\mathrm{F}^\mathrm{f} = 9.5$\,mg/l, and $f = 1.05/41$. %, and $L = 0.105$\,m.
    As in Figure~\ref{notTW}, the black dashed line marks the transition in time discretisation: times before this line are spaced logarithmically, while times after are spaced linearly.
    }
    \label{TMRC_only_longitudinal}
\end{figure}

Finally, Figure~\ref{TMRC_only_longitudinal} shows the predicted fluoride concentration profiles within the filter over time, using the reduced model. In contrast to the full model, which exhibits two distinct timescales, the reduced model solution evolves as a near-travelling wave at large times. This behaviour is expected, as the omission of the MRC components eliminates the secondary timescale associated with slower adsorption dynamics.

\section{Discussion and Conclusions}

{In the present study, we carefully derived the chemically based models for fluoride adsorption using mineral-rich carbon (MRC), chemically treated mineral-rich carbon (TMRC), and their mixture.} These models were validated against new experimental data, alongside experimental data originally published in~\cite{auton2023mathematical}. We first calibrated the individual MRC and TMRC models using batch isotherm and kinetic experiments, then extended these models to describe fixed-bed column filters. The resulting composite column model relies on four parameters that have been determined from the MRC and TMRC isotherms, leaving only three fitting parameters for the breakthrough curves.

This chemically based model showed strong predictive capabilities, particularly in reproducing breakthrough curves across varying inlet fluoride concentrations using a constant set of reaction rates, {which shows that the model correctly captures the dominant chemical mechanisms  occuring during the removal process}. The fitting resulted in a coefficient of determination (R$^2$) of over 0.991 in all cases and  dimensionless sum of square errors (SSE)   below 0.0632 for all curves. This shows that the model is robust and responds as expected to varying control conditions. This is indicative that the model is successfully captures the dominant removal mechanisms found in the filter.  

By developing models separately for MRC and TMRC and utilising them with the individual MRC and TMRC isotherm curves, we were able to significantly reduce the parameter fitting burden for the column model. Moreover, we proposed a simplified, reduced model that considers only ion-exchange resulting from the TMRC fraction. Despite MRC outweighing the TMRC by approximately 40 times, the TMRC fraction was found to be predominately responsible for the fluoride removal.  This reduced model, requiring just one fitting parameter for the breakthrough curves, still achieved excellent agreement with experimental data, with an R$^2$ greater than 0.983 for all curves and a dimensionless SSE of less than 0.116. The simplicity of this model means that it is straightforward and inexpensive to work with numerically. 

Although TMRC is a much more effective adsorbent than MRC ($\sim 10$ times), its  higher manufacturing cost and smaller grain size pose practical challenges, such as increased risk of clogging as the ratio of TMRC to MRC increases. Therefore, optimal filter design requires a careful balance between the content of MRC versus TMRC. 
Future work should explore the influence of varying  additional operational parameters in column filters, such as filter length, cross-sectional area and ratio of MRC:TMRC, on system performance. The models presented here offer a promising foundation for future efforts aimed at predicting filter lifespan and optimising filter design. This is of  particularly  importance for the filters currently being deployed in resource-limited settings such as rural West Bengal. Other approaches aimed at extending filter lifespan such as periodic agitation or mixing of the filter bed should also be evaluated.

Finally, further modelling and experimental studies should address the competitive adsorption between OH$^-$ and F$^-$, enabling a more complete understanding of pH-dependent effects. Incorporating the influence of multiple ionic species and examining a broader pH range would help refine the mechanistic understanding of fluoride adsorption, thereby enhancing the applicability and reliability of these models in real-world settings.

\section*{Data}
\noindent All \verb|MATLAB| codes used within this manuscript can be found at: \\  \verb|https://github.com/LucyAuton/Fluoride_removal_filters.git|

\section*{Acknowledgements}
This publication is part of the research projects PID2023-146332OB-C21 financed by  MCIN/ AEI/ 10.13039/501100011033/, and by “ERDF A way of making Europe” and the CERCA Programme of the Generalitat de Catalunya. The work was also supported by the Spanish State Research Agency, through the Severo Ochoa and Maria de Maeztu Program for Centres and Units of Excellence in R\&D (CEX2020-001084-M). AV is a Serra-Hunter fellow from the Serra-Hunter Programme of the Generalitat de Catalunya, and he acknowledges prior support from the Margarita Salas UPC postdoctoral grants funded by the Spanish Ministry of Universities with European Union funds - NextGenerationEU (UNI/551/2021 UP2021-034). Further, the authors acknowledge helpful discussions with Prof. Ian M. Griffiths of the University of Oxford, and Marc Martínez i Àvila's contributions to the initial stages of the project.

\section*{Author contributions: $\mathrm{\textbf{CRediT}}$}
\noindent\textbf{LCA}: Conceptualization; Formal analysis; Investigation; Methodology; Software; Supervision; Validation; Visualization; Writing – original draft; Writing – review and editing.\\
\textbf{SSR}: Data curation; Investigation; Methodology; Visualization; Writing – review and editing.\\
\textbf{SD}: Funding acquisition; Project administration; Resources; Supervision; Writing – review and editing.\\
\textbf{TGM}: Conceptualization; Funding acquisition; Investigation; Methodology; Project administration; Resources; Supervision; Writing – review and editing. \\
\textbf{AV}: Conceptualization; Data curation; Formal analysis; Investigation; Methodology; Software; Supervision; Validation; Writing – original draft; Writing – review and editing.

\clearpage 
\bibliographystyle{plainnat} 
\bibliography{E_refs}

\begin{thebibliography}{60}
\providecommand{\natexlab}[1]{#1}
\providecommand{\url}[1]{\texttt{#1}}
\expandafter\ifx\csname urlstyle\endcsname\relax
  \providecommand{\doi}[1]{doi: #1}\else
  \providecommand{\doi}{doi: \begingroup \urlstyle{rm}\Url}\fi

\bibitem[Aguareles et~al.(2023)Aguareles, Barrab{\'e}s, Myers, and
  Valverde]{aguareles2023mathematical}
M.~Aguareles, E.~Barrab{\'e}s, T.~G. Myers, and A.~Valverde.
\newblock Mathematical analysis of a {S}ips-based model for column adsorption.
\newblock \emph{Physica D: Nonlinear Phenomena}, 448:\penalty0 133690, 2023.

\bibitem[Alkurdi et~al.(2019)Alkurdi, Al-Juboori, Bundschuh, and
  Hamawand]{alkurdi2019bone}
S.~S.~A Alkurdi, R.~A. Al-Juboori, J.~Bundschuh, and I.~Hamawand.
\newblock Bone char as a green sorbent for removing health threatening fluoride
  from drinking water.
\newblock \emph{Environment International}, 127:\penalty0 704--719, 2019.

\bibitem[Auton et~al.(2023)Auton, \`{A}vila, Ravuru, De, Myers, and
  Valverde]{auton2023mathematical}
L.~C. Auton, M.~Mart\'{i}nez~I \`{A}vila, S.~S. Ravuru, S.~De, T.~G. Myers, and
  A.~Valverde.
\newblock Mathematical model for fluoride-removal filters.
\newblock \emph{Proceedings of the 22$^\mathrm{nd}$ ECMI conference, Wroclaw},
  2023.

\bibitem[Balasooriya et~al.(2022)Balasooriya, Chen, Korale~Gedara, Han, and
  Wickramaratne]{balasooriya2022applications}
I.~L. Balasooriya, J.~Chen, S.~M. Korale~Gedara, Y.~Han, and M.~N.
  Wickramaratne.
\newblock Applications of nano hydroxyapatite as adsorbents: a review.
\newblock \emph{Nanomaterials}, 12:\penalty0 2324, 2022.

\bibitem[Barathi et~al.(2014)Barathi, Kumar, Kumar, and
  Rajesh]{barathi2014graphene}
M.~Barathi, A.~S.~K. Kumar, C.~U. Kumar, and N~Rajesh.
\newblock Graphene oxide--aluminium oxyhydroxide interaction and its
  application for the effective adsorption of fluoride.
\newblock \emph{{RSC} Advances}, 4:\penalty0 53711--53721, 2014.

\bibitem[B{\'e}n{\'e}zeth et~al.(2008)B{\'e}n{\'e}zeth, Palmer, and
  Wesolowski]{benezeth2008dissolution}
P.~B{\'e}n{\'e}zeth, D.~A Palmer, and D.~J. Wesolowski.
\newblock Dissolution/precipitation kinetics of boehmite and gibbsite:
  Application of a p{H}-relaxation technique to study near-equilibrium rates.
\newblock \emph{Geochimica et Cosmochimica Acta}, 72:\penalty0 2429--2453,
  2008.

\bibitem[Bharti et~al.(2017)Bharti, Giri, and Kumar]{bharti2017fluoride}
V.~K. Bharti, A.~Giri, and K.~Kumar.
\newblock Fluoride sources, toxicity and its amelioration: a review.
\newblock \emph{Annals of Environmental Science and Toxicology}, 2\penalty0
  (1):\penalty0 021--032, 2017.

\bibitem[Bhatnagar et~al.(2011)Bhatnagar, Kumar, and
  Sillanp{\"a}{\"a}]{bhatnagar2011fluoride}
A.~Bhatnagar, E.~Kumar, and M.~Sillanp{\"a}{\"a}.
\newblock Fluoride removal from water by adsorption --- {A} review.
\newblock \emph{Chemical Engineering Journal}, 171:\penalty0 811--840, 2011.

\bibitem[Biswas et~al.(2007)Biswas, Saha, and Ghosh]{biswas2007adsorption}
K.~Biswas, S.~K. Saha, and U.~C. Ghosh.
\newblock Adsorption of fluoride from aqueous solution by a synthetic iron
  ({III})- aluminum ({III}) mixed oxide.
\newblock \emph{Industrial \& Engineering Chemistry Research}, 46:\penalty0
  5346--5356, 2007.

\bibitem[Cattarin et~al.(2009)Cattarin, Guerriero, Musiani, Tuissi, and
  V{\'a}zquez-G{\'o}mez]{cattarin2009electrochemical}
S.~Cattarin, P.~Guerriero, M.~Musiani, A.~Tuissi, and L.~V{\'a}zquez-G{\'o}mez.
\newblock Electrochemical etching of {NiTi} alloy in a neutral fluoride
  solution.
\newblock \emph{Journal of The Electrochemical Society}, 156\penalty0
  (12):\penalty0 {C}428--{C}434, 2009.

\bibitem[Chang(2009)]{chang2009chemistry}
R.~Chang.
\newblock \emph{Chemistry}.
\newblock McGraw-Hill, 10 edition, {J}an 2009.

\bibitem[Chatterjee(2018)]{chatterjee2018removal}
S.~Chatterjee.
\newblock \emph{Removal of Fluoride and Other Contaminants from Drinking
  Water}.
\newblock PhD thesis, IIT Kharagpur, 2018.

\bibitem[Chatterjee et~al.(2018{\natexlab{a}})Chatterjee, Jha, and
  De]{chatterjee2018novel}
S.~Chatterjee, S.~Jha, and S.~De.
\newblock Novel carbonized bone meal for defluoridation of groundwater: Batch
  and column study.
\newblock \emph{Journal of Environmental Science and Health, Part A},
  53\penalty0 (9):\penalty0 832--846, 2018{\natexlab{a}}.

\bibitem[Chatterjee et~al.(2018{\natexlab{b}})Chatterjee, Mukherjee, and
  De]{chatterjee2018defluoridation}
S.~Chatterjee, M.~Mukherjee, and S.~De.
\newblock Defluoridation using novel chemically treated carbonized bone meal:
  batch and dynamic performance with scale-up studies.
\newblock \emph{Environmental Science and Pollution Research}, 25:\penalty0
  18161--18178, 2018{\natexlab{b}}.

\bibitem[Chen et~al.(2022)Chen, Yang, Zhang, and Wu]{chen2022removal}
J.~Chen, R.~Yang, Z.~Zhang, and D.~Wu.
\newblock Removal of fluoride from water using aluminum hydroxide-loaded
  zeolite synthesized from coal fly ash.
\newblock \emph{Journal of Hazardous Materials}, 421:\penalty0 126817, 2022.

\bibitem[Cheng et~al.(2007)Cheng, Chalmers, and Sheldon]{cheng2007adding}
K.~K. Cheng, I.~Chalmers, and T.~A. Sheldon.
\newblock Adding fluoride to water supplies.
\newblock \emph{BMJ}, 335:\penalty0 699--702, 2007.

\bibitem[Craig et~al.(2015)Craig, Stillings, Decker, and
  Thomas]{craig2015comparing}
L.~Craig, L.~L. Stillings, D.~L. Decker, and J.~M. Thomas.
\newblock Comparing activated alumina with indigenous laterite and bauxite as
  potential sorbents for removing fluoride from drinking water in {G}hana.
\newblock \emph{Applied Geochemistry}, 56:\penalty0 50--66, 2015.

\bibitem[Craig et~al.(2017)Craig, Stillings, and Decker]{craig2017assessing}
L.~Craig, L.~L. Stillings, and D.~L. Decker.
\newblock Assessing changes in the physico-chemical properties and fluoride
  adsorption capacity of activated alumina under varied conditions.
\newblock \emph{Applied Geochemistry}, 76:\penalty0 112--123, 2017.

\bibitem[Do(1998)]{do1998adsorption}
D.~D. Do.
\newblock \emph{Adsorption analysis: {E}quilibria and kinetics}, volume~2 of
  \emph{Series on Chemical Engineering}.
\newblock Imperial College Press, 1998.

\bibitem[Gai et~al.(2022)Gai, Zhang, Yang, Sun, Jia, and
  Deng]{gai2022defluoridation}
W.-Z. Gai, S.-H. Zhang, Y.~Yang, K.~Sun, H.~Jia, and Z.-Y. Deng.
\newblock Defluoridation performance comparison of aluminum hydroxides with
  different crystalline phases.
\newblock \emph{Water Supply}, 22\penalty0 (4):\penalty0 3673--3684, 2022.

\bibitem[Ghorai and Pant(2004)]{ghorai2004investigations}
S.~Ghorai and K.~K. Pant.
\newblock Investigations on the column performance of fluoride adsorption by
  activated alumina in a fixed-bed.
\newblock \emph{Chemical Engineering Journal}, 98:\penalty0 165--173, 2004.

\bibitem[Gong et~al.(2012{\natexlab{a}})Gong, Qu, Liu, and
  Lan]{gong2012adsorption}
W.-X. Gong, J.-H. Qu, R.-P. Liu, and H.-C. Lan.
\newblock Adsorption of fluoride onto different types of aluminas.
\newblock \emph{Chemical Engineering Journal}, 189--190:\penalty0 126--133,
  2012{\natexlab{a}}.

\bibitem[Gong et~al.(2012{\natexlab{b}})Gong, Qu, Liu, and Lan]{gong2012effect}
W.-X. Gong, J.-H. Qu, R.-P. Liu, and H.-C. Lan.
\newblock Effect of aluminum fluoride complexation on fluoride removal by
  coagulation.
\newblock \emph{Colloids and {S}urfaces {A}: {P}hysicochemical and
  {E}ngineering {A}spects}, 395:\penalty0 88--93, 2012{\natexlab{b}}.

\bibitem[Hao and Huang(1986)]{hao1986adsorption}
O.~J. Hao and C.~P. Huang.
\newblock Adsorption characteristics of fluoride onto hydrous alumina.
\newblock \emph{Journal of Environmental Engineering}, 112:\penalty0
  1054--1069, 1986.

\bibitem[Hem and Roberson(1967)]{hem1967form}
J.~D. Hem and C.~E. Roberson.
\newblock \emph{Form and stability of aluminum hydroxide complexes in dilute
  solution}.
\newblock Chemistry of aluminium in natural water. US Government Printing
  Office, 1967.

\bibitem[Huang et~al.(2023)Huang, Zhang, Wang, Li, Zhang, Yang, He, and
  Gao]{huang2023enhanced}
S.~Huang, X.~Zhang, L.~Wang, D.~Li, C.~Zhang, L.~Yang, Q.~He, and B.~Gao.
\newblock Enhanced water defluoridation using ion channel modified
  hydroxyapatite: {E}xperimental, mechanisms and {DFT} calculation.
\newblock \emph{Applied Surface Science}, 615:\penalty0 156351, 2023.

\bibitem[Huyen et~al.(2023)Huyen, Phat, Tien, Thu, and Thoai]{huyen2023bone}
D.~T. Huyen, L.~N. Phat, D.~X. Tien, D.~P.~G. Thu, and D.~Q. Thoai.
\newblock Bone-char from various food-waste: {S}ynthesis, characterization, and
  removal of fluoride in groundwater.
\newblock \emph{Environmental Technology \& Innovation}, 32:\penalty0 103342,
  2023.

\bibitem[Inglezakis et~al.(2020)Inglezakis, Balsamo, and
  Montagnaro]{inglezakis2020liquid}
V.~J. Inglezakis, M.~Balsamo, and F.~Montagnaro.
\newblock Liquid--solid mass transfer in adsorption systems---an overlooked
  resistance?
\newblock \emph{Industrial \& Engineering Chemistry Research}, 59:\penalty0
  22007--22016, 2020.

\bibitem[Jia et~al.(2015)Jia, Zhu, Sun, Luo, Yu, Kong, and
  Liu]{jia2015fluoride}
Y.~Jia, Z.~Zhu, B.-S.and~Jin, B.~Sun, T.~Luo, X.-Y. Yu, L.-T. Kong, and J.-H.
  Liu.
\newblock Fluoride removal mechanism of bayerite/boehmite nanocomposites: roles
  of the surface hydroxyl groups and the nitrate anions.
\newblock \emph{Journal of Colloid and Interface Science}, 440:\penalty0
  60--67, 2015.

\bibitem[Kaminsky et~al.(1990)Kaminsky, Mahoney, Leach, Melius, and
  Jo~Miller]{kaminsky1990fluoride}
L.~S. Kaminsky, M.~C. Mahoney, J.~Leach, J.~Melius, and M.~Jo~Miller.
\newblock Fluoride: benefits and risks of exposure.
\newblock \emph{Critical Reviews in Oral Biology \& Medicine}, 1\penalty0
  (4):\penalty0 261--281, 1990.

\bibitem[Levenspiel(1999)]{levenspiel1998chemical}
O.~Levenspiel.
\newblock \emph{Chemical {R}eaction {E}ngineering}.
\newblock John {W}iley \& {S}ons, third edition, 1999.

\bibitem[Lin et~al.(2020)Lin, Chen, Hong, Huang, and Huang]{lin2020role}
J.-Y. Lin, Y.-L. Chen, X.-Y. Hong, C.~Huang, and C.~P. Huang.
\newblock The role of fluoroaluminate complexes on the adsorption of fluoride
  onto hydrous alumina in aqueous solutions.
\newblock \emph{Journal of Colloid and Interface Science}, 561:\penalty0
  275--286, 2020.

\bibitem[Locatelli and Schoen(1999)]{locatelli1999random}
M.~Locatelli and F.~Schoen.
\newblock Random linkage: a family of acceptance/rejection algorithms for
  global optimisation.
\newblock \emph{Mathematical Programming}, 85:\penalty0 379--396, 1999.

\bibitem[Medellin-Castillo et~al.(2014)Medellin-Castillo, Leyva-Ramos,
  Ocampo~Perez, Flores-Cano, and Berber-Mendoza]{medellin2014adsorption}
N.~A. Medellin-Castillo, E.~Leyva-Ramos, R .and Padilla-Ortega,
  R.~Ocampo~Perez, J.~V. Flores-Cano, and M.~S. Berber-Mendoza.
\newblock Adsorption capacity of bone char for removing fluoride from water
  solution. {R}ole of hydroxyapatite content, adsorption mechanism and
  competing anions.
\newblock \emph{Journal of Industrial and Engineering Chemistry}, 20\penalty0
  (6):\penalty0 4014--4021, 2014.

\bibitem[Misic et~al.(1982)Misic, Sudo, M., and Kawazoe]{misic1982liquid}
D.~M. Misic, Y.~Sudo, Suzuki M., and K.~Kawazoe.
\newblock Liquid-to-particle mass transfer in a stirred batch adsorption tank
  with non-linear isotherm.
\newblock \emph{Journal of Chemical Engineering of Japan}, 15\penalty0
  (1):\penalty0 67--70, 1982.

\bibitem[Mosiman(2021)]{mosiman2021probing}
D.~S. Mosiman.
\newblock \emph{Probing structure-property relationships of calcium
  hydroxyapatite defluoridation to enhance performance}.
\newblock PhD thesis, University of Illinois at Urbana-Champaign, 2021.

\bibitem[Mosiman et~al.(2021)Mosiman, Sutrisno, Fu, and
  Mari{\~n}as]{mosiman2021internalization}
D.~S. Mosiman, A.~Sutrisno, R.~Fu, and B.~J. Mari{\~n}as.
\newblock Internalization of fluoride in hydroxyapatite nanoparticles.
\newblock \emph{Environmental Science \& Technology}, 55:\penalty0 2639--2651,
  2021.

\bibitem[Myers(2024)]{myers2024time}
T.~G. Myers.
\newblock Is it time to move on from the {B}ohart-{A}dams model for column
  adsorption?
\newblock \emph{International Communications in Heat and Mass Transfer},
  159:\penalty0 108062, 2024.

\bibitem[Myers et~al.(2020)Myers, Font, and Hennessy]{Myer20b}
T.~G. Myers, F.~Font, and M.~G. Hennessy.
\newblock Mathematical modelling of carbon capture in a packed column by
  adsorption.
\newblock \emph{Applied Energy}, 278:\penalty0 115565, 2020.
\newblock \doi{10.1016/j.apenergy.2020.115565}.

\bibitem[Myers et~al.(2023)Myers, Cabrera-Codony, and
  Valverde]{myers2023development}
T.~G. Myers, A.~Cabrera-Codony, and A.~Valverde.
\newblock On the development of a consistent mathematical model for adsorption
  in a packed column (and why standard models fail).
\newblock \emph{International Journal of Heat and Mass Transfer}, 202:\penalty0
  123660, 2023.

\bibitem[Nie et~al.(2012)Nie, Hu, and Kong]{nie2012enhanced}
Y.~Nie, C.~Hu, and C.~Kong.
\newblock Enhanced fluoride adsorption using {Al (III)} modified calcium
  hydroxyapatite.
\newblock \emph{Journal of Hazardous Materials}, 233--234:\penalty0 194--199,
  2012.

\bibitem[Nordin et~al.(1999)Nordin, Sullivan, Phillips, and
  Casey]{nordin1999mechanisms}
J.~P. Nordin, D.~J. Sullivan, B.~L. Phillips, and W.~H. Casey.
\newblock Mechanisms for fluoride-promoted dissolution of bayerite
  [$\beta$-{Al(OH)}$_3$(s)] and boehmite [$\gamma$-{AlOOH}]: $^{19}${F-NMR}
  spectroscopy and aqueous surface chemistry.
\newblock \emph{Geochimica et Cosmochimica Acta}, 63\penalty0 (21):\penalty0
  3513--3524, 1999.

\bibitem[Ozsvath(2009)]{ozsvath2009fluoride}
D.~L. Ozsvath.
\newblock Fluoride and environmental health: A review.
\newblock \emph{Reviews in Environmental Science and Biotechnology},
  8:\penalty0 59--79, 2009.

\bibitem[Patmonoaji et~al.(2023)Patmonoaji, Tahta, Tuasikal, She, Hu, and
  Suekane]{patmonoaji2023dissolution}
A.~Patmonoaji, M.~A. Tahta, J.~A. Tuasikal, Y.~She, Y.~Hu, and T.~Suekane.
\newblock Dissolution mass transfer of trapped gases in porous media: {A}
  correlation of {S}herwood, {R}eynolds, and {S}chmidt numbers.
\newblock \emph{International Journal of Heat and Mass Transfer}, 205:\penalty0
  123860, 2023.

\bibitem[Pearson(1959)]{pearson1959note}
J.~R.~A. Pearson.
\newblock A note on the \lq\lq{D}anckwerts\rq\rq \ boundary conditions for
  continuous flow reactors.
\newblock \emph{Chemical Engineering Science}, 10:\penalty0 281--284, 1959.

\bibitem[Ren et~al.(2019)Ren, Yu, Phillips, Wang, Ji, Pan, and
  Li]{ren2019molecular}
C.~Ren, Z.~Yu, B.~L. Phillips, H.~Wang, J.~Ji, B.~Pan, and W.~Li.
\newblock Molecular-scale investigation of fluoride sorption mechanism by
  nanosized hydroxyapatite using $^{19}${F} solid-state {NMR} spectroscopy.
\newblock \emph{Journal of Colloid and Interface Science}, 557:\penalty0
  357--366, 2019.

\bibitem[Russo et~al.(2015)Russo, Tesser, Trifuoggi, Giugni, and
  Di~Serio]{russo2015dynamic}
V.~Russo, R.~Tesser, M.~Trifuoggi, M.~Giugni, and M.~Di~Serio.
\newblock A dynamic intraparticle model for fluid--solid adsorption kinetics.
\newblock \emph{Computers and Chemical Engineering}, 74:\penalty0 66--74, 2015.

\bibitem[Russo et~al.(2024)Russo, D'Angelo, Salvi, Paparo, Fortunato,
  Cepollaro, Tarallo, Trifuoggi, Di~Serio, and Tesser]{russo2024fluoride}
V.~Russo, A.~D'Angelo, C.~Salvi, R.~Paparo, M.~E. Fortunato, E.~M. Cepollaro,
  O.~Tarallo, M.~Trifuoggi, M.~Di~Serio, and R.~Tesser.
\newblock Fluoride adsorption on hydroxyapatite: {F}rom batch to continuous
  operation.
\newblock \emph{Journal of Environmental Chemical Engineering}, 12:\penalty0
  111973, 2024.

\bibitem[Sarkar et~al.(2006)Sarkar, Banerjee, Pramanick, and
  Sarkar]{sarkar2006use}
M.~Sarkar, A.~Banerjee, P.~P. Pramanick, and A.~R. Sarkar.
\newblock Use of laterite for the removal of fluoride from contaminated
  drinking water.
\newblock \emph{Journal of Colloid and Interface Science}, 302:\penalty0
  432--441, 2006.

\bibitem[Sternitzke et~al.(2012)Sternitzke, Kaegi, Audinot, Lewin, Hering, and
  Johnson]{sternitzke2012uptake}
V.~Sternitzke, R.~Kaegi, J.-N. Audinot, E.~Lewin, J.~G. Hering, and C.~A.
  Johnson.
\newblock Uptake of fluoride from aqueous solution on nano-sized
  hydroxyapatite: examination of a fluoridated surface layer.
\newblock \emph{Environmental Science \& Technology}, 46:\penalty0 802--809,
  2012.

\bibitem[Sundaram et~al.(2008)Sundaram, Viswanathan, and
  Meenakshi]{sundaram2008defluoridation}
C.~S. Sundaram, N.~Viswanathan, and S.~Meenakshi.
\newblock Defluoridation chemistry of synthetic hydroxyapatite at nano scale:
  {E}quilibrium and kinetic studies.
\newblock \emph{Journal of Hazardous Materials}, 155:\penalty0 206--215, 2008.

\bibitem[Tomar et~al.(2015)Tomar, Thareja, and Sarkar]{tomar2015enhanced}
G~Tomar, A~Thareja, and S.~Sarkar.
\newblock Enhanced fluoride removal by hydroxyapatite-modified activated
  alumina.
\newblock \emph{International Journal of Environmental Science and Technology},
  12:\penalty0 2809--2818, 2015.

\bibitem[Tripathy et~al.(2006)Tripathy, Bersillon, and
  Gopal]{tripathy2006removal}
S.~S. Tripathy, J.-L. Bersillon, and K.~Gopal.
\newblock Removal of fluoride from drinking water by adsorption onto
  alum-impregnated activated alumina.
\newblock \emph{Separation and Purification Technology}, 50:\penalty0 310--317,
  2006.

\bibitem[Tung and Skrtic(2001)]{tung2001interfacial}
M.~S Tung and D.~Skrtic.
\newblock Interfacial properties of hydroxyapatite, fluoroapatite and
  octacalcium phosphate.
\newblock \emph{Monographs in Oral Science}, 18 (Octacalcium
  Phosphate):\penalty0 112--129, 2001.

\bibitem[Ugray et~al.(2007)Ugray, Lasdon, Plummer, Glover, Kelly, and
  Mart\'{i}]{ugray2007scatter}
Z.~Ugray, L.~Lasdon, J.~Plummer, F.~Glover, J.~Kelly, and R.~Mart\'{i}.
\newblock Scatter search and local {NLP} solvers: {A} multistart framework for
  global optimization.
\newblock \emph{INFORMS Journal on Computing}, 19\penalty0 (3):\penalty0
  328--340, 2007.

\bibitem[Vithanage et~al.(2012)Vithanage, Jayarathna, Rajapaksha, Dissanayake,
  Bootharaju, and Pradeep]{vithanage2012modeling}
M.~Vithanage, L.~Jayarathna, A.~U. Rajapaksha, C.~B. Dissanayake, M.~S.
  Bootharaju, and T.~Pradeep.
\newblock Modeling sorption of fluoride on to iron rich laterite.
\newblock \emph{Colloids and Surfaces A: Physicochemical and Engineering
  Aspects}, 398:\penalty0 69--75, 2012.

\bibitem[Wakao and Funazkri(1978)]{wakao1978effect}
N.~Wakao and T.~Funazkri.
\newblock Effect of fluid dispersion coefficients on particle-to-fluid mass
  transfer coefficients in packed beds: Correlation of {S}herwood numbers.
\newblock \emph{Chemical Engineering Science}, 33:\penalty0 1375--1384, 1978.

\bibitem[Wendimu et~al.(2017)Wendimu, Zewge, and
  Mulugeta]{wendimu2017aluminium}
G.~Wendimu, F.~Zewge, and E.~Mulugeta.
\newblock Aluminium-iron-amended activated bamboo charcoal ({AIAABC}) for
  fluoride removal from aqueous solutions.
\newblock \emph{Journal of Water Process Engineering}, 16:\penalty0 123--131,
  2017.

\bibitem[{(WHO)}(2019)]{world2019preventing}
World Health~Organisation {(WHO)}.
\newblock Preventing disease through healthy environments. {I}nadequate or
  excess fluoride: a major public health concern.
\newblock {WHO} publication: Chemical Safety and Health Unit,
  WHO/CED/PHE/EPE/19.4.5, 2019.

\bibitem[Yang et~al.(2007)Yang, Sun, Wang, and Forsling]{yang2007surface}
X.~Yang, Z.~Sun, D.~Wang, and W.~Forsling.
\newblock Surface acid--base properties and hydration/dehydration mechanisms of
  aluminum (hydr)oxides.
\newblock \emph{Journal of Colloid and Interface Science}, 308:\penalty0
  395--404, 2007.

\end{thebibliography}

\end{document}